\documentstyle[11pt,aaspp4,flushrt]{article}

\begin{document}

\title{The Absolute Magnitude and Kinematics of RR Lyrae Stars via 
Statistical Parallax}
\author{Andrew C. Layden}
\affil{Department of Physics \& Astronomy, \\McMaster University,
Hamilton, Ontario, L8S 4M1, Canada\\$layden@physics.mcmaster.ca$}
\authoremail{alayden@physics.mcmaster.ca}
\author{Robert B. Hanson}
\affil{University of California Observatories / Lick Observatory, \\
University of California, Santa Cruz, CA 95064\\$hanson@ucolick.org$}
\authoremail{hanson@ucolick.org}
\author{Suzanne L. Hawley\altaffilmark{1}}
\affil{Department of Physics \& Astronomy, Michigan State University,  
E. Lansing, MI 48824\\$slh@pillan.pa.msu.edu$}
\authoremail{slh@pillan.pa.msu.edu}
\author{Arnold R. Klemola}
\affil{University of California Observatories / Lick Observatory,\\
University of California, Santa Cruz, CA 95064\\$klemola@ucolick.org$}
\authoremail{klemola@ucolick.org}
\author{Christopher J. Hanley}
\affil{Department of Physics \& Astronomy, Michigan State University,  
E. Lansing, MI 48824\\$chris@pillan.pa.msu.edu$}
\authoremail{chris@pillan.pa.msu.edu}

\altaffiltext{1}{NSF Young Investigator}

\lefthead{Layden {\em et al.}}
\righthead{$M_V(RR)$ via Statistical Parallax}

\begin{abstract}

We present new statistical parallax solutions for the absolute
magnitude and kinematics of RR Lyrae stars.  We have combined new
proper motions from the Lick Northern Proper Motion program with new
radial velocity and abundance measures to produce a data set that is
50\% larger, and of higher quality, than the data sets employed by
previous analyses.  Based on an {\it a priori} kinematic study, we
separated the stars into halo and thick disk sub-populations.  We
performed statistical parallax solutions on these sub-samples, and
found $M_V(RR) = +0.71 \pm 0.12$ at $\langle$[Fe/H]$\rangle$ = --1.61
for the halo (162 stars), and $M_V(RR) = +0.79 \pm 0.30$ at
$\langle$[Fe/H]$\rangle$ = --0.76 for the thick disk (51 stars).  The
solutions yielded a solar motion $\langle V \rangle = -210 \pm 12$
km~s$^{-1}$ and velocity ellipsoid $(\sigma_U, \sigma_V, \sigma_W) =
(168 \pm 13, 102 \pm 8, 97 \pm 7)$ km~s$^{-1}$ for the halo.  The
values were $\langle V \rangle = -48 \pm 9$ km~s$^{-1}$ and
$(\sigma_U, \sigma_V, \sigma_W) = (56 \pm 8, 51 \pm 8, 31 \pm 5)$
km~s$^{-1}$ for the thick disk.  Both are in good agreement with
estimates of the halo and thick disk kinematics derived from both RR
Lyrae stars and other stellar tracers.  Monte Carlo simulations
indicated that the solutions are accurate, and that the errors may be
smaller than the estimates above.  The simulations revealed a small
bias in the disk solutions, and appropriate corrections were derived.
The large uncertainty in the disk $M_V(RR)$ prevents ascertaining the
slope of the $M_V(RR)$--[Fe/H] relation.  Using a zero point defined
by our halo solution and adopting a slope of 0.15 mag~dex$^{-1}$, we
find that (1) the distance to the Galactic Center is $7.6 \pm 0.4$
kpc; (2) the mean age of the 17 oldest Galactic globular clusters is
$16.5 _{-1.9}^{+2.1}$ Gyr; and (3) the distance modulus of the LMC is
$18.28 \pm 0.13$ mag.  Estimates of $H_0$ which are based on an LMC
distance modulus of 18.50 (e.g., Cepheid studies) increase by 10\% if
they are recalibrated to match our LMC distance modulus.

\end{abstract}


\section{Introduction}

The absolute magnitude of the RR Lyrae variables, $M_V(RR)$, is
integral to determining distances to old stellar systems in our Galaxy
and to other nearby galaxies.  For example, RR Lyraes are widely used
to measure the distances to Galactic globular clusters, to the
Galactic Center, and to many members of the Local Group.  In addition,
the distances to individual field RR Lyrae stars in the thick disk and
halo of our Galaxy enable us to determine the kinematic and spatial
distributions of these populations.

Precise distances to globular clusters are also necessary to determine
their ages (e.g., Buonanno {\em et al.} 1989).
The variation of $M_V(RR)$ with abundance, [Fe/H],
strongly affects the derived age spread and age-metallicity relation
of the Galactic globular cluster system.  These quantities in turn
place strong constraints on scenarios describing the formation of the
Galaxy, specifying the rate of halo formation and whether the
chemical enrichment of the halo proceeded in a uniform, global fashion,
(Eggen {\em et al.}  1962) or within autonomous star-forming fragments
(Searle \& Zinn 1978).  The zero-point of the $M_V(RR)$--[Fe/H]
relation sets the mean absolute age of the globular cluster system,
thus placing a critical lower limit on the age of the Universe.

The RR Lyraes in a given globular cluster are observed to have a very
narrow range of intensity-mean magnitudes, typically $\sigma_V$ =
0.06--0.15 mag (Sandage 1990a).  Though the $M_V(RR)$ variation from
cluster to cluster over a broad range in [Fe/H] is not so well
understood, it is clear that RR Lyrae stars have the potential to be
excellent standard candles.

Historically, there has been discussion over the slope of the
$M_V(RR)$--[Fe/H] relation, with values ranging between $\Delta
M_V/\Delta$[Fe/H] = 0 to 0.4 mag dex$^{-1}$ (corresponding to a
globular cluster age range of $\gtrsim$5 Gyr to approximately zero).
Recently, a consensus has begun to form that the slope has a value of
0.15--0.20 (e.g., Carney {\em et al.} 1992 and Chaboyer 1995, though
see Sandage 1993 and Mazzitelli {\em et al.} 1995 for dissenting
opinions).

However, the zero-point of the relationship continues to defy a
consensus.  At the characteristic abundance of the halo, [Fe/H] =
--1.6, $M_V(RR)$ values range from 0.45 to 0.75 mag, and usually fall
either toward the brighter or the fainter end of this range.  
Chaboyer (1995) showed that this ``two value'' effect translates to a
$\sim$22\% difference in the derived ages of the Galactic globular
clusters, and in fact represents the dominant uncertainty in the
determination of cluster ages.

A number of methods have been used to estimate $M_V(RR)$, including
Baade-Wesselink (surface brightness) analyses (Jones {\em et al.}
1992), main sequence fitting of globular clusters (Buonanno {\em et
al.} 1989), application of stellar pulsation theory to field stars
(Sandage 1990b), horizontal branch evolution theory (Lee {\em et al.}
1990), calibrating the LMC RR Lyraes using other LMC distance
estimates (Walker 1992, Gould 1995), and the statistical parallax
method (Hawley {\em et al.} 1986, Strugnell {\em et al.}  1986).


The essence of the statistical parallax (``stat-$\pi$'') method is to
balance the radial-velocity-derived kinematics of a homogeneous
stellar sample with its kinematics as derived from proper motions.
The former are independent of distance, while the latter are
distance-dependent.  They are balanced through a simultaneous solution
for a distance scale factor.
Hawley {\em et al.} (1986) discussed the stat-$\pi$ solutions
performed prior to 1985, and described the shortcomings in the methods
that were employed.  They argued that only a complete treatment using
a maximum-likelihood formulation, together with a minimization
technique which is tolerant of inter-dependent variables, can produce
accurate solutions.

Two modern studies have employed these techniques (Hawley {\em et al.}
1986, and Strugnell {\em et al.} 1986, hereafter referred to as HJBW
and SRM, respectively).  HJBW compiled proper motion, radial velocity,
apparent magnitude, and abundance data from the literature to produce
a sample of $\sim$140 stars.  SRM employed the virtually same set of
data.  The two groups obtained similar values for $M_V(RR)$, the only
difference being in the details of the adopted reddenings and
determination of the apparent magnitudes.  Both groups concluded that
the sample was too small to constrain the slope of the
$M_V(RR)$--[Fe/H] relation.

Since then, considerable new data have become available, indicating
that it is an opportune time for a new stat-$\pi$ analysis.  The Lick
Northern Proper Motion (NPM) program (Klemola, Jones \& Hanson 1987)
has measured absolute proper motions for over 1000 RR Lyrae stars.
These data are more uniform and have a more accurate zero-point than
the proper motions employed in previous analyses, mainly because the
plates, obtained from a single telescope and covering most of the
Northern hemisphere, were measured and reduced onto a single inertial
frame tied to external galaxies.  Meanwhile, Layden (1994) determined
abundances and radial velocities for over 300 RR Lyraes, including
most of those in the HJBW and SRM studies.  Blanco (1992) showed that
the abundances used in those studies were of variable accuracy and
zero-point.  Layden's [Fe/H] measures are on a self-consistent system
and are typically accurate to 0.15--0.20 dex.  Thus, a new stat-$\pi$
solution will reap the benefits of a 50\% larger sample size (213
stars) and higher quality data.

Furthermore, Layden (1995) showed that the local RR Lyrae sample
breaks fairly cleanly into thick disk and halo populations at [Fe/H] =
--1.  The existence of two kinematically distinct populations had not
been recognized in previous stat-$\pi$ studies, in part because the
existing $\Delta S$ abundances were of insufficient accuracy to
provide the the required abundance resolution.  As a result, the two
populations had been mixed.  One worries that such mixing might have
resulted in the $M_V(RR)$--[Fe/H] slope being under-estimated.  In the
worst case, population mixing might lead to errant solutions, since
the stat-$\pi$ method is a simultaneous solution for kinematics and
$M_V(RR)$.
We consider it safest to treat the disk and halo separately in our
solutions.

This paper reports the findings of our new stat-$\pi$ solutions, based
on the improvements described above.  Throughout the paper, we employ
the stat-$\pi$ code used by HJBW.  In Sec. 2, we describe in detail
the data used in our stat-$\pi$ solutions.  In Sec. 3, we begin our
analysis with an inverted approach; we assume an $M_V(RR)$--[Fe/H]
relation, and compute the distance and three space velocity components
of each RR Lyrae star in our sample.  This gives us insight into how
best to divide the sample during the stat-$\pi$ solutions.  In Sec. 4,
we present Monte Carlo simulations which enable us to investigate the
accuracy of our solutions, and to search for any inherent biases
produced by the stat-$\pi$ technique.  In Sec. 5, we present the
absolute magnitude and kinematic results of the stat-$\pi$ solutions
for the observed stars.  In Sec. 6, we compare our $M_V(RR)$ results
with those obtained by other authors using other techniques.  In
Sec. 7, we discuss the implications of our results for some basic
properties of the Galaxy and the Universe.  We close with a short
summary of our findings.

\section{Data}


\subsection{Proper Motions}

Our primary source of proper motions for this study is the Lick
Northern Proper Motion (NPM) program (Klemola, Jones \& Hanson 1987).
The NPM program is a photographic survey measuring precise absolute
proper motions, on an inertial system defined by 50,000 faint galaxies
($16 \lesssim B \lesssim 18$), for over 300,000 stars with $8 \lesssim
B \lesssim 18$, covering the northern two-thirds of the sky ($\delta >
-23^{\circ}$), based on plates taken between 1947 and 1988 with the Lick
51~cm Carnegie Double Astrograph.  Details of the NPM observing, plate
measurement, and reduction procedures are given by Klemola {\it
et~al.} (1987).

Part~I of the NPM program, covering the 72\% of the northern sky lying
outside the heavily obscured regions of the Milky Way, was completed
in 1993, with the release of the Lick NPM1 Catalog (Klemola, Hanson \&
Jones 1993; Hanson 1993), containing 149,000 stars.  The stellar
content of the NPM1 Catalog is detailed in the NPM1
Cross-Identifications (Klemola, Hanson \& Jones 1994a; Hanson \&
Klemola 1994).  Comprehensive error analyses (Klemola, Hanson \& Jones
1994b) have determined the RMS error of the NPM absolute proper
motions to be $\sigma_\mu = 0\farcs 5\ {\rm cent}^{-1}$ in each
coordinate, corresponding to a transverse velocity error $\sim 25\
{\rm km~sec^{-1}~kpc}^{-1}$.  The NPM1 Catalog contains over 1000 RR
Lyrae variables (Klemola {\it et~al.}  1994a, Appendix 2).  Some 300
of these have $B < 14$ (corresponding to $D \lesssim 4$ kpc) and may
be individually useful for stat-$\pi$ studies.

The particular value of the NPM proper motions for this work is that
they are on an absolute reference frame.  By contrast, previous RR
Lyrae stat-$\pi$ solutions have relied on relative proper motions,
measured with respect to field stars (generally with $10 \lesssim B
\lesssim 12$), and corrected to absolute by assuming the net motions
of the reference stars from models of Galactic kinematics and
rotation.  Using relative proper motions inevitably raises the
possibility that the resulting kinematics (and luminosities) may
depend to some extent on the input assumptions.

Specifically, the HJBW and SRM stat-$\pi$ solutions used the list of
proper motions for 168 RR Lyraes compiled by Wan, Mao \& Ji (1980;
hereafter referred to as WMJ).  WMJ added new relative proper motions
from the Shanghai Observatory (Wan, He, Zhu \& Li 1979) and other
sources to the previous such compilation by Hemenway (1975).

Because of their absolute character, we adopted the NPM proper motions
as the primary source for our stat-$\pi$ database.  Our search of the
NPM1 Catalog, using the NPM1 Cross-Identifications, found 171 RR Lyrae
stars from Layden's (1994) list.  However, the NPM1 Catalog does not
cover low Galactic latitudes ($|b| \lesssim 10^{\circ}$), nor the southern
sky below $-23^{\circ}$ declination.  To attain the complete sky coverage
needed for reliable stat-$\pi$ solutions, we used the WMJ compilation
as a secondary source where NPM proper motions were not available,
adding another 42 stars.  Using the WMJ data raises two practical
problems:

First, should the WMJ motions be corrected to the NPM absolute system?
The heterogeneous nature of the WMJ compilation might make any single
correction doubtful.  However, Clube \& Dawe (1980b) suggested that
the existing RR Lyrae proper motions needed a large correction ``in
the direction of Galactic rotation'' $\Delta \mu = -1\farcs 4 \pm
0\farcs 8 \ {\rm cent}^{-1}$ due to incorrect motions of the reference
stars, making $M_V(RR)\ \sim 0.05$ to 0.1~mag fainter.  So large a
systematic error should be easily detectable by comparing the NPM and
WMJ proper motions.

Second, how should the WMJ data be weighted relative to the NPM
motions?  WMJ estimated the errors for each star from the
repeatability of the existing proper motions for that star, but the
number of determinations is generally small, so these estimates may
not be individually reliable.  For example, the range of
$\sigma_\mu({\sc wmj})$ is very large, some values are incredibly
small (one is zero!), and the correlation between each star's right
ascension and declination proper motion errors is very poor.
One-fourth of the WMJ stars have $\sigma_\mu$ in one coordinate more
than three times the value in the other coordinate.  Because all the
sources cited in WMJ used methods which should produce errors of equal
size in either coordinate, this is plainly an artifact of the WMJ
error analysis.  However, the overall mean and RMS errors ($0\farcs49$
and $0\farcs68 \ {\rm cent}^{-1}$ in $\mu_\alpha$; $0\farcs43$ and
$0\farcs61 \ {\rm cent}^{-1}$ in $\mu_\delta$) are quite comparable to
values given in the literature (Wan {\it et~al.} 1979; Hemenway 1975),
and may in fact be reliable error estimates.

To answer these questions, we compared data for the 109 stars in
common between the NPM1 Catalog and the WMJ compilation.  The
comparison was done twice; using $\Delta \mu = \mu({\sc npm}) -
\mu({\sc wmj})$ in equatorial coordinates ($\alpha, \delta$) and in
Galactic coordinates ($l,b$).  Normal probability plots (Lutz \&
Hanson 1992) were used for robust estimates of the mean differences
and RMS dispersions.  Plots of $\Delta\mu$ vs.  $\mu,\ \alpha,\
\delta,\ l,\ b$ and magnitude were examined for any dependence on
these observational variables.  Finally, we assessed the significance
of the WMJ individual proper motion errors $\sigma_\mu$({\sc wmj}).
Two principal results were found, bearing on each of the questions
posed above.

First, the WMJ proper motions in Galactic longitude have a small but
significant mean difference with NPM1.  We found
$$ <\Delta \mu_l> \ = \ -0\farcs 23 \pm 0\farcs 08 \ {\rm cent}^{-1} $$
$$ <\Delta \mu_b> \ = \ -0\farcs 03 \pm 0\farcs 08 \ {\rm cent}^{-1}.$$
As Clube \& Dawe (1980b) suggest, such a systematic difference may reflect 
erroneous reference star motions, but our result is six times smaller
than theirs.
Consequently, any effects on $M_V(RR)$ would be at the 0.01 mag level.
So, there is no need to correct the WMJ data for our stat-$\pi$ solutions.

Second, the WMJ proper motion errors are only useful as an overall
average, not on a star-by-star basis.  Figure 1 shows the NPM--WMJ
proper motion differences versus the WMJ listed errors.  In each
coordinate, the stars with the smaller errors ($\sigma_\mu\ \sim
0\farcs 2 \ {\rm cent}^{-1}$) scatter just as much as the stars with
the larger errors ($\sigma_\mu\ > 0\farcs 4 \ {\rm cent}^{-1}$).  For
the very smallest errors ($\sigma_\mu\ < 0\farcs 1 \ {\rm cent}^{-1}$)
the scatter is smaller, but the offsets from zero are large.  A
constant error describes the data much better than does the large
variation in $\sigma_\mu({\sc wmj})$.


Figure 2 quantifies this, testing whether the RMS error actually
increases with $\sigma_\mu$.  In each coordinate, the 109 stars were
sorted by the size of $\sigma_\mu({\sc wmj})$.  Figure 2 plots the
running values of the RMS nominal error and the actual RMS dispersion
of $\Delta\mu$ as a function of this rank.  The nominal error
$\sigma({\sc tot})$ is the quadratic sum of $\sigma_\mu({\sc wmj})$
and $0\farcs 5 \ {\rm cent}^{-1} = \sigma_\mu({\sc npm})$.  The
smooth, slowly rising curve labeled ``tot'' in Figure 2 is the running
RMS value of $\sigma({\sc tot})$.  If the WMJ errors are correct, then
$\sigma(\Delta\mu)$, the RMS dispersion of the proper motion
differences, should follow this curve.  The jagged line labeled
``$\Delta\mu$'' traces the running value of the actual RMS dispersion.
In each coordinate, $\sigma(\Delta\mu)$ quickly rises to a large value
($\sim 0 \farcs 85 \ {\rm cent}^{-1}$ in $\mu_\alpha$, $\sim 0 \farcs
80 \ {\rm cent}^{-1}$ in $\mu_\delta$) by rank $\sim 10$, and remains
at that level out to rank $\gtrsim 100$, rising somewhat at the end.
This proves that the real error of the WMJ proper motions is roughly
constant, independent of the listed error $\sigma_\mu({\sc wmj})$.
Quadratically subtracting $\sigma_\mu({\sc npm}) = 0\farcs 5 \ {\rm
cent}^{-1}$ from these values gives $\sigma_\mu({\sc wmj}) = (0\farcs
69,\ 0\farcs 62)\ {\rm cent}^{-1}$ in ($\alpha,\delta$).  Almost
identical error estimates were obtained from normal probability plots
of $\Delta\mu(\alpha,\delta)$.  These estimates agree almost perfectly
with the RMS values of the WMJ errors cited above.

In view of this result, we gave all WMJ proper motions equal weights
in our stat-$\pi$ solutions.  Test solutions with $\sigma_\mu({\sc
wmj}) = 0\farcs 5$ and $0\farcs 7 \ {\rm cent}^{-1}$, corresponding to
the mean and RMS WMJ errors, respectively, gave very similar results.
We decided to adopt the smaller value, $\sigma_\mu({\sc wmj}) =
0\farcs 5 \ {\rm cent}^{-1} = \sigma_\mu({\sc npm})$, in order to
weight all areas of the sky equally in the stat-$\pi$ solutions
(Section 5).

\subsection{Abundances}

Our primary source for RR Lyrae metal abundances is the work of Layden
(1994, hereafter referred to as L94).  These abundances are based on
the relative strengths of the \ion{Ca}{2} K line and the Balmer lines
H$\delta$ through H$\beta$, analogous to the $\Delta$$S$ abundance
technique of Preston (1959).  The abundance scale is tied to the
[Fe/H] abundance scale for globular clusters developed by Zinn \& West
(1984), and the individual [Fe/H] values are typically accurate to
0.15--0.20 dex.  Lambert {\em et al}.  (1995) have measured new
high-dispersion abundances for a number of the stars in L94, and find
excellent agreement with those results.  Jurcsik \& Kovacs (1996) also
discuss the high quality of the L94 abundances.

Previous statistical parallax solutions have used $\Delta$$S$ values as
the metallicity indicator.  Since then, Blanco (1992) has shown that
$\Delta$$S$ values available to those authors were of variable accuracy
and zero-point.  The [Fe/H] sample of L94 is both self-consistent and
contains many more stars than a sample of $\Delta$$S$ values collected
from the literature.

However, there are a few bright RR Lyraes in the literature which are
not included in the list of L94.  For these, we adopt the literature
$\Delta$$S$ values, converted to [Fe/H] using Eqn. 6 of L94.  We note
that some of these values, those taken from Hemenway (1975), are
actually {\it inferred} from photoelectric indices or a
period-amplitude-$\Delta$$S$ relation.  We discuss these stars further
in Sec. 2.6.

\subsection{Radial Velocities}

L94 also measured radial velocities for the stars in his
sample, and combined them with literature velocities to produce a
catalog of radial velocities, the most accurate currently available,
for over 300 nearby RR Lyrae stars.  L94 is our primary source of
radial velocities.  Velocities for a few stars not observed by L94
were taken from the literature compilation shown in Table 1 of L94.
We include these, adopting for their errors the typical errors for
each source derived in Sec. 2 of L94.

\subsection{Apparent Magnitudes}

The existing photometry on field RR Lyrae stars is a surprisingly
heterogeneous data set.  There are three principal works in the Johnson
$V$-band.  The work of Sturch (1966) is comprised mainly of
observations at minimum light; that of Bookmyer {\em et al}. (1977), which
contains as a subset the better-known work of Fitch, Wisniewski \&
Johnson (1966), and is purported to be on the same photometric system;
and that of Clube \& Dawe (1980b, hereafter referred to as CD80).
Barnes \& Hawley (1986, hereafter referred to as BH86) show that the
photometric system of Sturch is in good agreement with that of CD80,
and that both are offset from the work of Fitch {\em et al.}.  We therefore
adopt the CD80 photometry as the standard to which we will compare
other photometric works.

Many of the existing RR Lyrae light curves have incomplete phase
coverage, so it is difficult to obtain their intensity-mean apparent
magnitudes with accuracy.  However, various relations exist in the
literature which allow this quantity to be calculated from light curve
extrema, rise times, etc. (e.g., Fitch {\em et al.} 1966, CD80).  BH86
recomputed the coefficients for two of these methods using modern,
self-consistent data, and find that the method of CD80 gives the tighter
relation.  We adopt this parameterization along with their
coefficients,
$$\langle V \rangle_I = V_{min} - 0.375~ \Delta V - 0.040,$$
where $\langle V \rangle_I$ is the intensity-mean magnitude,
$V_{min}$ is the magnitude at minimum light, and $\Delta V$ is the
light curve amplitude.  By using the CD80 photometry as the basis of
our photometric system, and by employing intensity-mean magnitudes
computed from the preceding equation, our photometry system is
equivalent to that of BH86.

We computed $\langle V \rangle_I$ for the CD80 and Bookmyer {\em et
al}.  data sets, and performed a linear regression between them to
obtain a transformation between the two photometric data sets (see
Table 1, line 1).   We then adopted data values of
CD80 (57 stars) as the primary data, and supplemented it with the
values from Bookmyer {\em et al}. that had been transformed onto the
CD80 system (81 additional stars). 

The resulting data set was used to transform the Walraven photometry of
Lub (1977) onto the CD80 system.  The transformation, given in line 2
of Table 1, is in close agreement with the relation of Pel (1976).
This resulted in 7 additional stars being added to the data set.

This data set was then used to transform the CCD photometry of Schmidt
{\em et al} (1991, 1995) onto the CD80 system (see line 3 of Table 1),
adding 36 stars to the database.

Preliminary photometry from Layden (1996; 8 stars) was also included,
though no transformations were possible since there were no stars in
common with the database.  Data for 24 additional stars were adopted
from the photometric compilation of L94 (his Table 9), after
converting from the $\langle V \rangle_I$ definition of Fitch {\em et
al.} 1966 to that of CD80.

Clearly, this approach is not ideal, since it relies on the
statistical transformations between photometric systems, and assumes
that the CD80 system is equivalent to the modern systems used by
observers (e.g., Landolt 1992) and theorists (e.g., Lee {\em et al.}
1990).  The approach has the advantage of reducing systematic errors
by placing all the stars on the same photometric system.  Ultimately,
the transformations result in changes, typically --0.07 mag, which are
small compared to the 0.2--0.3 mag disagreements which arise between
different methods of measuring $M_V(RR)$.  Furthermore, the sense of
the transformation is to brighten the literature values; had we used
the fainter photometry, the $M_V(RR)$ values we derived would have
been fainter as well.  Obtaining self-consistent photometry at the
level of several hundredths of a magnitude is a problem shared, but
seldom mentioned, by all observers attempting to measure $M_V(RR)$.

\subsection{Interstellar Absorption}

SRM argued that reddenings derived from the Burstein \& Heiles (1982)
\ion{H}{1} reddening maps provide the most accurate and consistent
estimates of RR Lyrae reddenings, and hence absorption (we assume
$A_V/E(B-V)$ = 3.1).  We therefore use Burstein \& Heiles values when
they are available, i.e., for stars more than 10$^{\circ}$ from the
Galactic plane.  We reduce their tabulated reddenings by an amount
consistent with a uniform dust distribution with an exponential scale
height of 100 pc (see L94).  In most cases this is a small or
negligible correction.

For stars less than 10$^{\circ}$ from the plane, we adopt the
reddening values given by Blanco (1992), which are derived from the
stars' colors at minimum light.  When neither are available, we
interpolate between the Burstein \& Heiles reddening at
$\pm$10$^{\circ}$, and that of FitzGerald (1968, 1987) at
$\pm$0.5$^{\circ}$, at the longitude of the star.  The latter is
clearly a poor solution, but it is the best available until accurate
minimum-light colors can be obtained for these stars.  Fortunately,
it was used for only 9 stars.

\subsection{The Final Database}

Using the data sources described above, we find that 213 stars have
values for all five of the fundamental data types: proper motion,
abundance, radial velocity, apparent magnitude, and reddening.  This
is substantially more than were used in the recent studies of
HJBW (142 stars) and SRM (139 stars).  Note that we do not consider
Bailey type-$c$ RR Lyraes in this study, only type-$ab$ stars.

This database is presented in Table 2.  
The first column gives the variable star name, and the second column
gives the NPM1 catalog number.  Following this are the galactic
longitude and latitude (in degrees) and the adopted proper motions in
right ascension and declination (in arcsec cen$^{-1}$).  The seventh
and eighth columns give the adopted radial velocity and its error (in
km~s$^{-1}$).  Next is the adopted abundance, [Fe/H].  The tenth
column gives the adopted intensity-mean apparent $V$ magnitude, and
the eleventh column gives the adopted interstellar absorption.  The
twelfth column gives references for the sources of the proper motion,
abundance, photometry, and interstellar absorption, as listed at the
end of the table.  The final column indicates whether a star was
treated as a disk (1) or halo (0) star under the three disk/halo
definitions discussed in Sec. 3.  

NPM proper motions are used for 171 of the 213 stars in our sample.
Abundances from L94 are used for 187 of the stars, and apparent
magnitudes computed directly from $V$--band photometry are used for
182 of the stars.

Given the dominance of NPM proper motions in our catalog, one wonders
if the distribution of stars on the sky is skewed, and whether this
would introduce a bias into our stat-$\pi$ solutions (e.g., Croswell,
Latham \& Carney 1987).  Regarding the former, we find that 63\% of
our sample lies above the celestial equator; we are weighted to the
North celestial hemisphere, but not overwhelmingly so.  Similarly,
67\% of our stars lie North of the Galactic plane.  However, these
asymmetries can not produce the kind of biases discussed by Croswell
{\it et al.}, since our sample contains no proper motion bias.  As
Klemola {\it et al.} (1987) describe, samples of ``astrophysically
interesting'' stars, such as RR Lyraes, were selected for the NPM
program in advance of the plate measurements, and no star was omitted
from the NPM1 Catalog because its measured proper motion proved to be
small.  This type of pre-selection is true of our secondary proper
motion source as well.  As confirmation, we note that our data show no
sign of the ``Croswell effect.''  The number of stars moving away from
the Galactic plane is almost identical to the number moving toward it:
106 $vs.$ 108, respectively.

Finally, we note specifics of interest concerning several stars.  (1)
BX Dra was shown by Schmidt {\em et al.} (1995) to be an eclipsing
binary rather than an RR Lyrae (Kholopov 1985); it was removed from
our database.  (2) L94 found SV Boo to have [Fe/H] = --0.43 from a
single low-quality spectrum, whereas Hemenway (1975) quoted
$\Delta$$S$ = 7 ([Fe/H]$_{L94}$ = --1.55).  Since the kinematics of SV
Boo suggest it belongs to the halo, we adopt the Hemenway abundance.
(3) The radial velocity of BB Pup was revised to 98 $\pm$ 9
km~s$^{-1}$ from that in L94 by eliminating the outlier velocity 255
$\pm$ 16 km~s$^{-1}$ from the list in Table 2 of L94.  (4) For three
stars, AE Dra, BD Dra, and BK Eri, the proper motion was improved by
removing one discordant measurement from the average value quoted in
the NPM1 Catalog.  (5) The seven stars listed with the abundance
source ``3'' in Table 2 all had photometrically-determined [Fe/H]
values from Hemenway (1975; see Sec. 2.2) which suggested that they
were thick disk stars.  However, their kinematics suggested that all
seven stars are halo members.  We therefore set [Fe/H] = --1.5 for
these stars.  (6) We found that RX CVn historically has been
misidentified.  The proper motion of WMJ and the radial velocity of
Joy (1950) give a space velocity greater than the escape velocity of
the Galaxy.  The NPM proper motion and the radial velocity measured by
L94 (see our Table 2) give a reasonable space velocity for a halo
star.  Inspection of the NPM plates shows that the measured star has a
nearby companion, probably a foreground dwarf, which was probably
mistakenly observed by WMJ and Joy.  The RR Lyrae is the eastern-most
of the pair.


\section{Kinematics and Population Separation}

Layden (1995, hereafter referred to as L95) showed that the RR Lyraes
separate into a halo and a (primarily thick) disk population at [Fe/H]
= --1.0.  We wish to see if this separation persists using our
improved database.

To do this, we computed provisional distances to the stars in our
sample using the $M_V(RR)$--[Fe/H] relationship of Carney, Storm \&
Jones (1992, hereafter referred to as CSJ), $M_V(RR)$ = 0.15[Fe/H] +
1.01 mag.  We then computed the stars' $U$, $V$, $W$ space velocities,
following Johnson \& Soderblom (1987) with one exception:  we take
$U$ as positive outward.
At distances of 1--2 kpc, typical of the distances in our sample, the
Sun-oriented $U$, $V$, $W$ frame can be misaligned by a small angle in
the $U$, $V$ plane ($RMS \approx 6$ deg) from the cylindrical Galactic
directions ($\pi$, $\theta$, $z$) at each star.  So, we rotated the
$U$, $V$, $W$ velocities into the $\pi$, $\theta$, $z$ frame, after
adding to $V$ the IAU standard rotational velocity $\Theta_0 = 220$ km
s$^{-1}$ (Kerr \& Lynden-Bell 1986), and after adding the
``dynamical'' solar motion (--9, +12, +7 km s$^{-1}$; Mihalas \&
Binney 1981, p.400) to correct $U$, $V$, $W$ to the Local Standard of
Rest (LSR).  The results are the velocity components $V_{\pi}$,
$V_{\theta}$, and $V_{z}$, where $V_{\pi}$ increases outwards from the
axis of Galactic rotation, $V_{\theta}$ increases in the direction of
Galactic rotation, and $V_{z}$ increases toward the North Galactic
Pole.

Figure 3 shows $V_{\pi}$ and $V_{z}$ plotted against $V_{\theta}$.
Clearly, the stars with [Fe/H] $<$ --1.0 have large velocity
dispersions and little net Galactic rotation, typical of the halo.
Meanwhile, the stars with [Fe/H] $> -1.0$ are clustered around
$V_{\theta}$ $\approx 200$ km~s$^{-1}$.  Figure 4 shows $V_{\theta}$
as a function of abundance. 

While the distribution of stars in Figures 3 and 4 is consistent with
the first-order view of a disk/halo separation at [Fe/H] = --1.0,
there are four stars with $V_{\theta}$ $< 60$ km~s$^{-1}$ at [Fe/H] $>
-1.0$, whose extreme kinematics clearly mark them as members of the
halo.  Similarly, there may be an excess of stars with $V_{\theta}$ $>
80$ km~s$^{-1}$ at [Fe/H] $< -1.0$, which may belong to the
``metal-weak thick disk'' (MWTD; Morrison, Flynn \& Freeman 1990).
L95 discussed the presence of such stars in his sample of RR Lyraes.

It is not possible to assign these stars individually to the disk or
halo populations without ambiguity.  We therefore separate the disk
from the halo using three distinct definitions, and perform the
stat-$\pi$ solutions for each set of definitions, in order to test the
effects of the different definitions on the derived kinematics and
absolute magnitudes.

The three disk/halo separation definitions are summarized in Table 3.
All three definitions assign the four low-$V_{\theta}$ stars with
[Fe/H] $> -1.0$ to the halo.  The first definition, similar to that of
Nissen \& Schuster (1991), admits a small number of MWTD RR Lyraes,
primarily with $-1.3 <$ [Fe/H] $< -1.0$, in agreement with L95.  The
second assumes that no MWTD ([Fe/H] $< -1.0$) RR Lyraes exist.  The
third definition admits a larger population of MWTD RR Lyraes, which
reaches to [Fe/H] $\approx -1.6$, more along the lines of the
population of red giants described by Morrison {\em et al}.  (1990).
Two stars, AO Peg and FU Vir, fit one or more of the disk definitions,
yet clearly belong to the halo based on their extreme kinematics:
($V_{\pi}, V_{\theta}, V_z$) = (--212, +236, --207) and (--178, +249,
--93), respectively.  We have moved these stars into the corresponding
halo definitions, as noted in Table 3.


We note that the choice of $M_V(RR)$--[Fe/H] relations used to compute
the distances does not significantly affect the separation.  Only one
star crosses the sloping disk/halo line in Fig. 4 when we change from
the CSJ $M_V(RR)$--[Fe/H] relation to that advocated by Sandage
(1993).


\section{Monte Carlo Simulations}

\subsection{Testing the Stat-$\pi$ Algorithm}

In Sec. 5 we will present the stat-$\pi$ analysis of our data.  First,
however, we will test the HJBW stat-$\pi$ algorithm itself, using
synthetic data with known properties (positions, velocities,
$M_V(RR)$, etc.).  This step seems vital to ensure reliable results.
Specifically, we test: (1) how accurately the stat-$\pi$ solutions
reproduce the kinematics and luminosities of the input data; (2) how
reliable the error estimates are; (3) the sensitivity to the number of
stars and their distribution on the sky; (4) whether all the free
parameters in the HJBW stat-$\pi$ algorithm are necessary; (5) whether
the small misalignment (Sec. 3) between $U, V, W$ and $V_{\pi},
V_{\theta}, V_z$ has any significant effects; (6) whether the results
are biased by any input assumptions; and (7) whether any corrections
are necessary for bias in the results.

The HJBW algorithm uses a simplex optimization technique to maximize
the likelihood in Murray's (1983) kinematic model (also used by SRM).
There are 11 free parameters: the solar motion ($U,V,W$), the velocity
ellipsoid ($\sigma_U, \sigma_V, \sigma_W$) with three covariances
($\sigma_{UV}, \sigma_{UW}, \sigma_{VW}$) to allow an arbitrary
orientation, the distance scale parameter $k$ and its dispersion
$\sigma_k$.  The solution also returns an approximate standard error
estimate $\sigma_i$ for each parameter.  The solution returns an
absolute magnitude $M_V$ by differential correction to a starting
value $M_A$, which for convenience we fix at +1.0.  We refer the
reader to HJBW for full details of the stat-$\pi$ algorithm.  We made
two modifications to the HJBW procedure: (1) we fixed an error which
caused the large velocity dispersion errors listed in Table 2 of HJBW;
and (2) when necessary, we inspected the data to reject extreme
outliers and repeated the solutions.

\subsection{Simulated Data Sets}

To perform Monte Carlo tests we generated an ensemble of simulated
data sets as outlined in Table 4.  These were designed to
realistically simulate the halo and disk subsamples of our real data
(Sec. 3).    

For the halo and disk simulations H1 and D1, $N_{stars} = (165,\ 50)$
were randomly assigned spatial positions ($X,Y,Z$) and [Fe/H] values
from uniform distributions with appropriate limits.  $M_V$ values were
then computed from the $M_V(RR)$--[Fe/H] relation of CSJ, with a
Gaussian cosmic dispersion $\sigma_M = 0.2$ mag.  Space velocities
($V_{\pi}$, $V_{\theta}$, $V_z$) were generated from a Gaussian
velocity ellipsoid whose parameters are given in Table 4.  Each star's
right ascension, declination, proper motion, radial velocity, and
apparent magnitude were then calculated, with Gaussian observational
errors 0.5 arcsec cen$^{-1}$ and 20 km s$^{-1}$ added to the proper
motion components and the radial velocity, respectively.  For each
simulation (H1, D1) we created $N_{trials} = (5,\ 20)$ different data
sets; $N_{trials}$ was set larger for the disk simulations because
$N_{stars}$ is proportionately smaller.

As discussed in Sec. 2, the actual distribution of our RR Lyraes on
the sky is far from uniform.  To test whether this affects the
stat-$\pi$ results, we prepared alternate data sets (H2, D2) using the
observed sky positions, apparent magnitudes, and metallicities from
Table 2, with the Disk-1/Halo-1 separation of Table 3.  Then,
$N_{trials}$ sets of synthetic proper motions and radial velocities
were randomly generated as above.

For each data set (H1, H2, D1, D2), stat-$\pi$ solutions were
performed as outlined in column 6 of Table 4.  Multiple solution sets
tested particular parameters in the HJBW model.  To test the effect of
the HJBW distance scale dispersion parameter $\sigma_k$ we ran two
solutions for each halo data set, with $\sigma_k = (0.0,0.1)$ as
indicated in column 7 of Table 4.  These solutions will be discussed
in Sec. 4.6.  For the data sets H2 and D2 we ran solutions with and
without the velocity covariance parameters ($\sigma_{UV}, \sigma_{UW},
\sigma_{VW}$), as indicated in column 8 of Table 4.  The reasons for
this will be discussed in Sec. 4.7.  Fifteen additional H2 data sets
(row 5 of Table 4) were created in order to test with higher precision
the important solution sets which excluded the velocity covariance
parameters.  The significance of these solutions will be discussed in
Sec. 4.8.  Finally, we note that for a few of the disk data sets, the
solutions failed to converge ($N_{conv} < N_{trials}$).  This will be
discussed in Sec. 4.5.

\subsection{Coordinate Frame Misalignment}

Before discussing the stat-$\pi$ solutions, we need to deal with an
apparent inconsistency between the data simulations and the HJBW
solution method.  We generated our simulated velocities using velocity
ellipsoids oriented to the cylindrical coordinate frame ($V_{\pi}$,
$V_{\theta}$, $V_z$) which is physically the most appropriate frame
for Galactic velocities.  However, the HJBW algorithm uses the
($U,V,W$) rectangular coordinate frame to compute the solar motion and
stellar velocity dispersions.  In Sec. 3, we noted that the ($U, V,
W$) coordinates can be misaligned with ($V_{\pi}$, $V_{\theta}$,
$V_z$), by a small angle (RMS $\sim 6$ deg).  It is important to show
that any effects of this misalignment are too small to significantly
affect our results.

We can do this directly with the simulated data, because for every
star we know the space velocity in each coordinate frame.  This lets
us measure, for every data set, any differences (1) between $\langle
U,V,W \rangle$ and $\langle V_{\pi}, V_{\theta}, V_z \rangle$ and (2)
between $\sigma_{(U,V,W)}$ and $\sigma_{(\pi,\theta,z)}$.  In the
remainder of this paper, we refer to the values computed directly from
the simulated data as the ``true'' values for the data set.

For the halo (H1, H2), the principal effect is the projection of the
long axis of the velocity ellipsoid ($\sigma_{\pi}$) partly onto the
$V$ axis, increasing $\sigma_V$ at the expense of $\sigma_U$.
Quantitatively, $\sigma_U = \sigma_{\pi} - 0.4$ km s$^{-1}$, and
$\sigma_V = \sigma_{\theta} + 0.5$ km s$^{-1}$.  Clearly, these
effects are far too small to be of any concern here.

For the disk (D1, D2), the principal effect of coordinate misalignment
is the partial projection of the rotation vector $V_{\theta} = 200$ km
s$^{-1}$ onto the $U$ axis.  Distant stars in the direction of
Galactic rotation $(l,b) = (90,0)$ get a negative contribution to
their $U$ velocity; stars toward (270,0) get a positive contribution.
The net effect is that $\sigma_U = \sigma_{\pi} + 3$~km~s$^{-1}$.
Also, $\langle V \rangle$ is decreased by 1 km s$^{-1}$.  Again, these
effects are considerably smaller than the observational errors, and
can be neglected in practice.

\subsection{Monte Carlo Simulation Results}

Stat-$\pi$ solutions for each of the 25 halo and 40 disk data sets
were performed as outlined in Table 4.  The results for each set of
solutions are summarized in Table 5.  The left side of Table 5 gives
results for the $V$ component of the solar motion, the three velocity
dispersions ($\sigma_U,\sigma_V,\sigma_W$), and $M_V$.  The $\Delta$
values on the right side of Table 5 represent the differences
(solution $-$ ``true'') for each quantity.  Table 5 omits the $U$ and
$W$ solar motion components, as these were always equal to the true
values to $\lesssim 1$ km s$^{-1}$.


Each solution set in Table 5 lists two rows of results.  The first row
gives the mean of each quantity.  The second row gives two different
estimates of the uncertainty in the quantity: $\langle \sigma_i
\rangle$ (left side of Table 5) gives the mean of the internal error
estimates returned by the HJBW code for each parameter, while SD
(right side) gives the RMS dispersion about each mean difference.  The
latter external error estimates reflect how precisely the stat-$\pi$
program returns the ``true'' values.  The values in row 5 of Table 5
are more precise than those quoted for the other halo simulations,
since they are based on 20 rather than 5 data sets.

The major result of the Monte Carlo simulations is that the HJBW
algorithm does an excellent job of returning the ``true'' input
parameters, to within a few km s$^{-1}$ for the velocities and
dispersions, and to within $\sim 0.1$ mag for $M_V$.  This is true for
both the halo and the disk simulations, and for both the random and
real space distributions.  Furthermore, in nearly all cases the
external errors (SD's of the $\Delta$'s) are no larger than the HJBW
program's internal error estimates $\langle \sigma_i \rangle$, and in
some cases they are considerably smaller.

Detailed examination of Table 5 shows some small systematic effects
which are worth considering further.  Note that 7 out of 8 $\langle
\Delta V \rangle$ values are positive, all 24 $\langle \Delta \sigma
\rangle$ values are negative, and 7 out of 8 $\langle \Delta M_V
\rangle$ values are positive.  Examination of the individual solutions
shows that these same effects occur on a solution-by-solution basis,
with $V$, $\sigma_{(U,V,W)}$, and $M_V$ all varying in tandem, linked
together by the HJBW distance scale parameter $k$.  The average values
are slightly biased toward a ``short'' distance scale ($k < 0,\ \Delta
M_V > 0$).  This bias is larger for the real space distribution data
sets (H2, D2) than for the random sets (H1, D1).  Whether we can or
should correct for this small bias will be discussed in Sec. 4.8.

For the halo (H) solutions, the internal error estimates $\langle
\sigma_i \rangle$ returned by the HJBW program tend to be larger than
the external errors.  For example, $\langle \sigma_i \rangle$ for
$M_V$ is 0.12 mag, while the SD of $\Delta M_V$ averages 0.08 mag.
For the kinematic parameters, 
$\langle \sigma_i\rangle$/SD $\gtrsim 2$.  
These results indicate
that the real accuracy of the halo solutions may be better than the
internal errors claim.  However, given the simplified nature of the
simulations, we will conservatively adopt the internal error estimates
in discussing our real data solutions (Sec. 5).

For the disk (D) solutions, the internal and external errors for $M_V$
generally agree, though there is a small discrepancy for the kinematic
parameters, $\langle \sigma_i\rangle$/SD $\approx 1.4$.  In addition,
the standard error of $M_V$ is $\sim$0.3 mag, 3--4 times as large as
for the halo.  The failure of the $M_V$ errors to follow an $N^{-1/2}$
relation may mean that $N_{stars} = 50$ is near the lower limit for
successful solutions (see Sec. 4.5), but it must also reflect the fact
that the stat-$\pi$ method inherently works better for a population
with a larger velocity dispersion.

The test solutions (H1.0, H2.0) with the HJBW distance scale
dispersion parameter $\sigma_k$ set to zero will be discussed in
Section 4.6.  The test solutions (H2d, D2d) without the velocity
ellipsoid covariances ($\sigma_{UV}, \sigma_{UW}, \sigma_{VW}$) will
be discussed in Section 4.7.

\subsection{Sample Size and Solution Convergence}

Since the number of stars in our disk samples (both real and synthetic
data) is near or below the lower limit (50 stars) that HJBW found
necessary for successful solutions, we need to ask how reliable our
disk solutions can be with so few stars.  The symptom of having too
few stars is that the HJBW likelihood function becomes
ill-conditioned, and the iterative solution fails to converge to
finite output parameters (hence $N_{conv} < N_{trials}$ in Table 4).

Surprisingly, given the HJBW result, over 90\% of our Monte Carlo disk
solutions, and all 3 real-data disk solutions (Sec. 5) did in fact
converge.  Moreover, Table 5 shows that the disk output parameters are
well-determined, i.e., they have little bias and accurate error
estimates.  The convergence of our Monte Carlo solutions for
$N_{stars} \leq 50$ may reflect the Gaussian nature of our
simulations, in contrast to the vagaries of the real data that HJBW
used.  In our experience, the HJBW algorithm is not resistant to large
outliers; for real data, this makes the disk/halo separation quite
critical.  The success of our real-data disk solutions for $N_{stars}
\simeq 40$ is most likely due to the better separation we achieved
using [Fe/H] and $V_\pi$ (Sec. 3, Fig. 4) instead of $\Delta S$ and
period (HJBW).

Without doing many more simulations, it is not possible for us to
state what the true lower limit on $N_{stars}$ may be.  Nor is this
necessary, since the clear result of our Monte Carlo simulations is
that solutions that do converge give reliable results.

\subsection{Effects of Cosmic Dispersion in $M_V(RR)$}

Both HJBW and SRM found a strong correlation in their stat-$\pi$
solutions between $M_V(RR)$ and $\sigma_M$, the cosmic dispersion in
$M_V(RR)$.  In the Murray (1983) model, the cosmic dispersion is
parameterized by $\sigma_k$, the dispersion in the distance scale
parameter $k$.  This correlation effectively prevents solving for
$\sigma_k$; instead this parameter must be fixed at a value chosen to
represent a reasonable value of $\sigma_M$.  Equation 8 of HJBW
relates $\sigma_M$, $\sigma_k$, and $k$.  For $\sigma_k = 0$,
$\sigma_M = 0$.  For $\sigma_k = 0.1$ and $\langle k \rangle = -0.1$
from our solutions, $\sigma_M \simeq 0.24$~mag.

Observations show this to be a reasonable range of $\sigma_M$.
Sandage (1990a) found the intrinsic dispersion of RR Lyrae magnitudes
within a globular cluster (i.e., at a single metallicity) to be
$\sigma_V$ = 0.06--0.15 mag.  For our field RR Lyraes there will be an
additional dispersion $\sigma_s$ proportional to the slope of the
$M_V(RR)$--[Fe/H] relation.  We estimated $\sigma_s$ numerically by
populating various $M_V(RR)$--[Fe/H] relations with ``stars'' having
the [Fe/H] distribution of our halo sample.  The total $\sigma_M$ is
then the quadratic sum of $\sigma_V$ and $\sigma_s$. For $\sigma_V =
0.1$~mag, using the CSJ $M_V(RR)$--[Fe/H] relation (slope =
+0.15~mag~dex$^{-1}$) we obtain $\sigma_M = 0.11$~mag.  Using the
steeper slope (+0.39) advocated by Sandage (1990b) gives $\sigma_{Mv}
= 0.17$~mag.  To reach $\sigma_{Mv} = 0.24$~mag, we must adopt an
extreme value of $\sigma_V = 0.2$~mag along with the Sandage (1990b)
slope.

Consequently, a reasonable range of $\sigma_k$ to test in our Monte
Carlo solutions is $0.0 \leq \sigma_k \leq 0.1$.  Solution sets (H1.0,
H1) and (H2.0, H2) with $\sigma_k = (0.0,\ 0.1)$ respectively (Table
4) let us evaluate the effects on $M_V(RR)$.  The results in Table 5
indicate that $M_V(RR)$ comes out $\sim$0.03 mag brighter for
$\sigma_k = 0.1$ than for $\sigma_k = 0.0$.  Similar results were
found by HJBW and SRM.  Table 5 shows that the assumed value of
$\sigma_k$ does not affect the derived kinematics.  Clearly, our
choice of $\sigma_k$ will only have a small effect on our stat-$\pi$
results.  To be conservative, we will adopt $\sigma_k = 0.1$ to
analyze our real data (Sec. 5).  For a given solution the true value
of $M_V(RR)$ thus may be a few hundredths of a magnitude fainter than
the value we derive.

\subsection{Velocity Ellipsoid Covariances}

Three of the 11 parameters in the Murray (1983) model are the
covariances ($\sigma_{UV}, \sigma_{UW}, \sigma_{VW})$ which allow the
velocity ellipsoid to have an arbitrary orientation with respect to
the principal Galactic directions.  The results of HJBW strongly
suggest that this may not be necessary in practice.  Given the
apparent benefits of eliminating unneeded parameters from the model
(especially for the disk, where $N_{stars} \lesssim 50$ does not
greatly exceed the number of free parameters), it seems wise to use
our Monte Carlo simulations to test whether the covariances
($\sigma_{UV}, \sigma_{UW}, \sigma_{VW}$) are needed in our real-data
stat-$\pi$ solutions.

Because our simulated data were generated with no correlations among
the ($U,V,W$) velocities, the covariances ($\sigma_{UV}, \sigma_{UW},
\sigma_{VW}$) returned by the stat-$\pi$ solutions simply reflect
random scatter in the data.  Thus the covariances for the real-data
solutions can be tested for statistical significance by comparison
with the Monte Carlo simulations.

To do this, for each stat-$\pi$ solution we calculated the correlation
coefficients (CC's) $\rho_{UV} = \sigma_{UV} / \sigma_U \sigma_V$,~
etc.  For each of the sets of simulated data listed in Table 4,
we computed the mean, SD, and range of the CC's of
the individual trials.  For all the simulations, the mean correlations
were near zero, as expected.  For the halo, (both H1 and H2) each of
the three SD's was $\lesssim 0.1$.  By comparison, the RMS value of
the real-data halo CC's was 0.10, with none of the CC's exceeding the
range of the simulated values.  For the disk, the SD's were $\sim 0.2$
for set D1 (random space distribution), and $\sim 0.3$ for set D2
(real space distribution).  The RMS value of the real-data disk CC's
was 0.13; again none of the CC's exceeded the range of the simulated
values.

From these tests we conclude that there is no evidence, for either
the halo or the disk RR Lyraes, that the velocity ellipsoid deviates
from the principal directions ($U,V,W$).  Consequently, we can run our
stat-$\pi$ solutions with the covariance parameters ($\sigma_{UV},
\sigma_{UW}, \sigma_{VW}$) removed from the model.  The entries in
Table 5 for solution sets H2d and D2d show the results of these
solutions on the data sets H2 and D2.  There is little difference in
the kinematics or absolute magnitudes produced by the new solutions,
save for a slight tendency for the errors to be reduced.  Analyzing
our real data (Sec. 5) without the unneeded covariances reduced the
disk $M_V$ errors by 10\%.

\subsection{Bias Corrections}

In Sec. 4.4 we found that the results in Table 5 suggested that the
stat-$\pi$ solutions may be slightly biased toward a ``short''
distance scale. The solution velocities are consistently several per
cent too small, and $M_V \lesssim 0.1$~mag too faint.  The bias was
largest ($\ \langle \Delta M_V \rangle = +0.13$~mag) for the data set
D2, the disk simulation with the space distribution of the real RR
Lyraes.  Although this bias is much less than the random error ($\sim
0.3$~mag) of a single disk solution, it may still be worth applying
corrections to our real-data results (Sec. 5).

The above-mentioned bias toward a ``short'' distance scale should not
be confused with a bias toward the {\em fainter} $M_V$ regime
discussed in Secs. 1, 6, and 7 (i.e., toward $M_V(RR) \approx +0.7$
mag at [Fe/H] $\approx -1.6$, rather than +0.4 mag).  It can easily be
demonstrated that our solutions have no
intrinsic bias toward any particular $M_V$ value; nor are the
solutions biased by the choice of the starting value $M_A$.  Tests
using simulated data sets based on $M_V(RR)$ 0.5 mag brighter than the
CSJ relation adopted in Sec. 4.2 correctly returned $M_V$ values 0.5
mag brighter, with unchanged kinematics.  These results were recovered
exactly when a brighter starting value $M_A = +0.5$ was used, and to
within $\sim0.01$ mag in $M_V$ and $\sim1$\% in the velocities even
when the usual $M_A = +1.0$ was used.

In the spirit of exploratory data analysis, we plotted $\Delta M_V$
versus the ``solution'' and ``true'' values of the velocities
($U,V,W)$ and dispersions $\sigma_{(U,V,W)}$ for all 65 simulations.
This led to the discovery that, for the real space distribution sets
(H2, D2), the bias was smallest when the velocity ellipsoid was
``long'', i.e. when $\sigma_U \gg \sigma_V$, and largest when the
velocity ellipsoid was ``round'', i.e. when $\sigma_U \lesssim
\sigma_V$.  Thus the disk is chiefly affected, while the halo is not.
The 15 additional H2d solutions outlined in Table 4 were performed to
illuminate this situation.

Figure 5 plots $\Delta M_V$ as a function of $\sigma_U(true) /
\sigma_V(true)$ for solution sets H2d and D2d.  Because of their
markedly different distributions on the sky, we consider the halo and
disk separately in Fig. 5.  For the halo, the small bias ($+0.036 \pm
0.019$ mag) is well-determined because of the increased number of H2d
solutions, but the slope is not statistically significant.  For the
disk, both the mean bias (+0.13 mag) and the slope (--0.69 mag) are
2-$\sigma$ significant.


We suggest that these results may be used to apply a bias correction
(subtracted from $M_V$) to our real-data results (Sec. 5) for the
disk, as a function of $\sigma_U / \sigma_V$.  (We choose not to apply
a $\Delta M_V$ correction to the halo results for two reasons.  The
halo correction would be small compared to the other sources of error
in $M_V(RR)$ discussed in Sec. 6; moreover the bias is compensated by
the roughly equal, but opposite, $\sigma_k$ effect discussed in
Sec. 4.6.)  For consistency, when we apply the disk $\Delta M_V$
correction in Sec. 5, we will also correct the disk kinematics for the
``short'' distance scale by enlarging the velocities and dispersions
by a factor of $10^{0.2~\Delta M_V}$.

Two objections may be raised to these corrections: first, that the
reason for the bias is not understood, and second, that the ``true''
value of $\sigma_U / \sigma_V$ is not known for the real data.  The
latter problem can be overcome by computing $\sigma_U$ and $\sigma_V$
directly from the data, as in Sec. 3.  Because we only need the ratio
$\sigma_U / \sigma_V$, any distance dependence from the assumed
$M_V(RR)$--[Fe/H] relation cancels out.

It remains mysterious to us why the disk stat-$\pi$ solutions are
slightly biased when the velocity ellipsoid is ``round''.  Since the
effect occurs for the real (but not a random) stellar distribution, it
must be caused by the the uneven distribution of the disk RR Lyraes on
the sky.  It might be possible to solve this puzzle with a much larger
set of simulations, but that is clearly beyond the scope of this
paper.  We note that maximum-likelihood methods in general are not
unbiased; Clube \& Dawe (1980a) found an equal $M_V$ bias ($-0.12$~mag)
in the opposite direction!  We conclude simply that since our disk
$M_V$ bias is relatively large and can be calibrated as a function of
the observational variables, we should apply it to our real-data
solutions.


\section{Stat-$\pi$ Solutions for Observed RR Lyrae Data}

Applying the lessons learned from our Monte Carlo simulations
(Sec. 4), we analyzed our real data (each of the Disk/Halo subsamples
in Table 3) by running the stat-$\pi$ program with $\sigma_k = 0.1$
and performing two sets of solutions, with and without the velocity
ellipsoid covariances ($\sigma_{UV}, \sigma_{UW}, \sigma_{VW}$).  As
in Sec. 4.7, the differences between the two sets of solutions were
small.  The kinematics generally changed by $<$ 1 km s$^{-1}$; $M_V$
averaged $\sim 0.02$~mag brighter in the solutions without the
covariance terms.  Most important, the errors returned by the
stat-$\pi$ program for the disk data sets were typically 10\% smaller
without the covariances, presumably owing to the larger number of
degrees of freedom attained by removing three free parameters from the
solutions.  Since the correlation coefficients ($\rho_{UV}, \rho_{UW},
\rho_{VW}$) did not prove to be significant (Sec. 4.7), we therefore
adopt these solutions, with the velocity ellipsoid aligned to the
principal Galactic directions ($U,V,W$), as our final results.

Table 6 presents, for each solution as discussed below, the solar
motion $\langle U,V,W \rangle$, the velocity ellipsoid
$\sigma_{(U,V,W)}$, and the absolute magnitudes.  Below each of these
entries is presented the standard error for that term, computed by the
stat-$\pi$ program.  Recall that the Monte Carlo simulations suggested
that the true uncertainties may be as much as 2 times {\em smaller}
than those quoted in the table.  The final column of Table 6 shows the
$M_V(RR)$ values after correction for the $\sigma_U/\sigma_V$ bias
discussed in Sec. 4.8.
 

\subsection{Kinematic Results}

Examination of Table 6 shows that the kinematics of the RR Lyraes do
not depend significantly on which disk/halo definition is employed.
Because Definition 3 of Table 3 gives the purest halo sample and the
largest, best-determined disk sample, we adopt Halo-3 and Disk-3 as
our best solutions in Table 6.  Thus, our best estimates of the net
rotation and velocity ellipsoid of the halo RR Lyraes are
$$\langle V \rangle = -210 \pm 12~{\rm km~s}^{-1},\ \ \ 
            V_{rot} =   22 \pm 12~{\rm km~s}^{-1}$$
$$(\sigma_U,\sigma_V,\sigma_W) = 
(168 \pm 13,\ 102 \pm 8,\ 97 \pm 7)~{\rm km~s}^{-1}.$$
For the disk RR Lyraes, after correcting (by a factor of 1.07) 
for the distance scale bias ($\Delta M_V = +0.15$ mag, Sec. 5.2), 
we obtain
$$\langle V \rangle =  -48 \pm  9~{\rm km~s}^{-1},\ \ \ 
            V_{rot} =  184 \pm  9~{\rm km~s}^{-1}$$
$$(\sigma_U,\sigma_V,\sigma_W) = ( 56 \pm 8,\ 51 \pm 8,\ 31 \pm 5)~{\rm km~s}^{-1}.$$ 
We note (see Sec. 4.4) that the HJBW stat-$\pi$ program implicitly
accounts for the effects of observational errors in the determination
of the velocity dispersion parameters, so these results are unbiased
estimates of the true velocity dispersions of the halo and disk RR
Lyrae populations.

These kinematic values are in excellent agreement with the RR Lyrae
kinematics derived by L95.  They also correspond quite well with the
kinematics of the thick disk and halo based on other tracer
populations ({\em cf.} Casertano {\em et al.} 1990; L95 Table 8), with
two possible exceptions.  First, the vertical velocity dispersion of
the disk, $\sigma_W = 31$ km~s$^{-1}$, is somewhat smaller than the
typically quoted value of 35--45 km~s$^{-1}$.  L95 suggested that the
RR Lyrae ``disk'' subsample contains stars from both the thick disk
and the old thin disk populations, such that the net kinematics are
intermediate between the two.  Unfortunately, the disk sample contains
too few stars for us to subdivide it and perform meaningful stat-$\pi$
solutions for separate thin and thick disk components.

Second, $\sigma_U$ for the halo RR Lyraes is large compared to many
other estimates, 168 km~s$^{-1}$ $vs.$ 120--155 km~s$^{-1}$.  The
cause of this effect is less clear.  It may be due to our removing
interloper thick disk stars more completely than other studies (L95),
or it may be related to a subtle selection bias experienced by RR
Lyraes.  For example, if the halo is composed of an accreted component
and a dissipatively-formed component (Zinn 1993, Majewski 1993), if
the components have different kinematics (e.g., Beers 1996), and if RR
Lyraes are more easily formed in one component than the other, then
using RR Lyraes as kinematic tracers would bias the kinematic results
to favor one or the other halo components, relative to their
representation in samples using other stellar tracers.  At present, it
seems that the halo RR Lyraes may be preferentially tracing the
accreted halo component, though a detailed analysis outside the scope of
this paper is required to further address this problem.

\subsection{Absolute Magnitudes}

In Table 6, $M_V$ is virtually the same for each of the three halo 
definitions.  As above, we adopt Halo-3 as our purest definition of 
the halo RR Lyraes.  This gives
$$M_V(RR) = +0.71 \pm 0.12~{\rm mag\ \ at\ \ \langle [Fe/H]} \rangle = -1.61.$$
As discussed in Secs. 4.8 and 6, no bias corrections have been applied
to the halo solutions.

For the disk, $M_V$ is more sensitive to which set of stars is used in
the stat-$\pi$ solution.  As in the Monte Carlo simulations (Sec 4.4),
the errors in the derived $M_{V}$'s for the disk are $\sim 3$ times
larger than for the halo.  Both effects are largely due to the
relatively small number of stars in the disk solutions.  Again we
adopt Disk-3 as the best solution.  Since the disk velocity ellipsoid
is quite ``round'' ($\sigma_U(true) / \sigma_V(true) = 1.06$), a bias
correction of $+0.15$~mag (Figure 5) was subtracted from the solution
value, giving
$$M_V(RR) = +0.79 \pm 0.30~{\rm mag\ \ at\ \ \langle [Fe/H]} \rangle = -0.76.$$

\subsection{Effects of Disk/Halo Separation}

To see what would have happened had we been unable to separate the
disk and halo RR Lyraes, we performed a stat-$\pi$ solution (last line
of Table 6) using all 213 stars.  This solution is comparable to
``Group RR~$ab$'' of HJBW (142 stars) and to ``Sample C$^{\prime}$''
(139 stars) of SRM.  Interestingly, the absolute magnitude ($+0.73 \pm
0.11$) for our ``All stars'' solution is almost exactly the same as
for the three halo solutions, but the kinematics are rather different.
For this mixture of disk and halo, $\langle V \rangle$, $\sigma_U$,
and $\sigma_W$ are smaller than for the pure halo, but $\sigma_V$ is
larger.  The velocity ellipsoid is much ``rounder''; $\sigma_U /
\sigma_V = 1.28$, vs.  1.65 for Halo-3.  HJBW and SRM found $\sigma_U
/ \sigma_V =$ 1.25 and 1.29, respectively, for the comparable groups.

These results point out that a good disk/halo separation is necessary
to get reliable kinematic results for the RR Lyraes.  The stat-$\pi$
method is robust enough to produce solutions for mixed populations
with distinctly different kinematics, successfully determining $M_V$,
but it derives kinematics not accurately representing either
population.

\subsection{$M_V(RR)$ Variations with [Fe/H]}

The fact that we have separate disk and halo solutions over a range of
almost 1 dex in [Fe/H] gives hope that we might obtain the slope of
the $M_V(RR)$--[Fe/H] relationship, a parameter of considerable
astrophysical importance (Sec. 1) which previous stat-$\pi$ studies
(e.g., HJBW, SRM) found difficult to determine.  Unfortunately, our
disk solutions are not sufficiently precise to meaningfully constrain
this slope.  Figure 6 depicts this fact graphically; the error bars on
the disk solutions easily admit slopes between 0 and +0.4
mag~dex$^{-1}$, the extreme values currently under debate.
Calculating the slope using the bias-corrected $M_V(RR)$ estimates for
Halo-3 and Disk-3 from Table 6, we obtain $\Delta M_V\ /\ \Delta{\rm
[Fe/H] = +0.09 \pm 0.38\ mag\ dex}^{-1}.$


To pursue the $M_V(RR)$--[Fe/H] slope further, we divided the Halo-1
sample at [Fe/H] = --1.55, giving equal-sized metal-rich and
metal-poor sub-groups.  We performed stat-$\pi$ solutions on each
group (Halo-1R and Halo-1P in Table 6, and the crosses in Figure 6).
These solutions show no indication of any slope within the halo.
Again, we are unable to constrain the slope of the $M_V(RR)$--[Fe/H]
relation within meaningful limits.


In a final effort to determine the $M_V(RR)$--[Fe/H] slope, we
attempted to incorporate this dependence directly into the stat-$\pi$
program by parameterizing the absolute magnitude term $M_A$ in Eqn. 5
of HJBW with with two coefficients $a$ and $b$, where
$$M_A\ =\ a ({\rm [Fe/H] - \langle [Fe/H]} \rangle)\ +\ b.$$ However,
preliminary solutions indicate that this makes the maximum-likelihood
solution ill-conditioned.  It may be possible to avoid this problem by
incorporating the [Fe/H] dependence into the distance scale correction
$k$ rather than $M_A$; we are continuing to work on this problem, and
any results will be reported in a future paper.


\section{Comparisons with other Measurements}

In Sec. 1, we mentioned the long history of efforts to determine
$M_V(RR)$.  CSJ provide an extensive review of much of this work,
which we shall not repeat here.  Table 7 presents the results of some
of these efforts \footnote{The $M_V(RR)$ values produced by the
stat-$\pi$ method are for ``typical'' RR Lyraes in the sample.  These
stars tend to be somewhat evolved off the Zero Age Horizontal Branch
(ZAHB).  Some $M_V(RR)$ estimates using different methods refer to
$M_V(RR)$ at the ZAHB.  We note these cases in Table 7, where we have
corrected $M_V(ZAHB)$ to $M_V(RR)$ using the Eqn. 4 of CSJ.}.  They
are plotted as a function of [Fe/H] in Figure 7 to facilitate
comparison with the results of our stat-$\pi$ solutions.  We refer the
interested reader to the individual references in Table 7 for more
detailed discussions on these methods.  

We begin by noting that our stat-$\pi$ solutions are in good agreement
with the results of the two recent applications of the method
($M_V(RR) = +0.68$, BH86; $M_V(RR) = +0.75$, SRM).  This is not a
complete surprise, given that the stars in those studies are all
included in the present study, albeit with improved data.  As shown in
Table 7 and Figure 7, the agreement between the various stat-$\pi$
solutions becomes even better when small corrections are made to bring
the previous results onto the system of reddenings and magnitudes used
in this paper. 
Like us, neither of the previous groups was able to detect a meaningful
trend in $M_V(RR)$ with [Fe/H] due to the small sample sizes.

At the characteristic abundance of the halo, [Fe/H] $\approx$ --1.6,
the various results shown in Figure 7 cover a range of 0.2--0.3 mag,
and appear to separate into a brighter and a fainter group.
Interestingly, whether a particular result is bright or faint does not
seem to be a function of the method employed.  For example, Buonanno
{\em et al.} (1990) found a bright zero-point by fitting 19 globulars to 5
subdwarfs, while CSJ found a faint zero-point by fitting a single
cluster to the subdwarf (of identical abundance) with the best
trigonometric parallax.  Similarly, using the Sandage period-shift
effect, Sandage (1990b) obtained a bright zero-point and a steep
slope.  Using data for a different set of field stars, adopting a
different effective temperature relation, and employing a different
mass-metallicity relation, CSJ obtained a moderate slope.  Fernley
(1993) used infra-red rather than optical observations of field and
cluster RR Lyraes in his period shift analysis, and found a bright
zero-point but a moderate slope.  Apparently, the current uncertainty
in $M_V(RR)$ is dominated by differences in the details of the methods
a particular author follows, and his or her choice of a particular data
set or reddening correction.  It is very difficult to determine
which of these assumptions are correct or incorrect at this level of
detail.

In deciding which methods shown in Figure 7 should be given the most
weight, it is worth noting several strengths of the stat-$\pi$ method.
First, stat-$\pi$ is independent of other distance determinations.  By
contrast, cluster main sequence fitting requires precise trigonometric
parallaxes to the nearby subdwarfs.  Calibrations of $M_V(RR)$ based
on LMC distances determined by other methods are similarly
complicated.  For example, the Cepheid calibration used by Walker
(1992) to obtain the LMC distance is based on main sequence fits of
Cepheid-bearing Galactic open clusters to the Pleiades, $and$ main
sequence fits of the Pleiades to local dwarfs with trigonometric
parallaxes.

A second strength of the stat-$\pi$ method is that it relies on a
simple, extremely well-tested model.  The kinematics of the Galaxy are
described by three mean velocities, three velocity dispersions, and
the orientation of this velocity ellipsoid relative to the cardinal
directions of the Galaxy.  Countless kinematic studies
over the past half century have shown this to be a very complete
description of local Galactic kinematics.  

By comparison, the models or basic assumptions on which many of the
other techniques rely are far more complex, and tend not to be so
well-tested.  For example, the Baade-Wesselink method depends on model
atmospheres to determine both the surface brightness constant, $S_o$,
and the observed-to-pulsation velocity correction, $p$ (Jones {\em et
al.} 1992).  Both the Baade-Wesselink and period shift methods rely on
a color-index $vs.$ effective temperature relation, and different
authors advocate different relations (Sandage 1990b, CSJ, Fernley
1993, Sandage 1993).  The period shift method also relies heavily on
the assumed RR Lyrae mass-metallicity relation, yet the mass estimates
from double-mode RR Lyraes are in serious conflict with those derived
from the Baade-Wesselink method, and to a lesser extent with masses
derived from HB theory (Fernley 1993, Yi {\it et al.} 1993).
Meanwhile, $M_V(RR)$ estimates from HB theory are dependent on the
color--temperature relation, the evolutionary models (including the
treatment of convection), and especially on the assumed main sequence
helium abundance, $Y_{MS}$ (Lee 1990).  $M_V(RR)$ values derived from
main sequence fits currently rely on theoretical isochrones to correct
the colors and/or magnitudes of the clusters and/or field subdwarfs to
a common metallicity (Buonanno {\it et al.} 1990, Bolte \& Hogan
1995).  The reader is referred to the papers noted above and in Table
7 for detailed discussion of these topics.

Our $M_V(RR)$ result for the halo is 0.06 mag brighter than the CSJ
value at [Fe/H] = --1.61, and agrees better with this and other
``faint'' $M_V(RR)$ values than the ``bright'' results shown in Fig. 7,
which are $\sim$0.2 mag brighter.
Given this dichotomy, it is valid to ask whether there are any
parameters we could change that would push our result to a brighter
value.  The results of SRM suggest that if we adopted the Sturch
(1966) reddening scale, based on the blanketing-corrected colors of RR
Lyraes at minimum light, we would obtain a result $\sim$0.11 mag
{\em brighter} than our current result.  However, SRM argue that the
Burstein \& Heiles (1982) \ion{H}{1} based redding scale is preferred.

In Sec. 2.4, we noted that the various sources of RR Lyrae photometry
suffer small inconsistencies in photometric standardization at the
0.07 mag level.  Had we used the un-transformed literature values, we
would have obtained an $M_V(RR)$ for the halo about 0.07 mag {\em
fainter}.

Several other effects could alter our $M_V(RR)$ result by very small
amounts.  In Sec. 4.6, we showed that observations constrain
$\sigma_k$ to be between 0.0 and 0.1.  By adopting $\sigma_k = 0.1$,
we obtained the brightest $M_V(RR)$ consistent with this constraint;
adopting a smaller value of $\sigma_k$ results in values of $M_V(RR)$
up to 0.03 mag {\em fainter}.  Had we corrected for the small bias
uncovered by our halo simulations (Sec. 4.8), our result would have
been $\sim$0.04 mag {\em brighter}.  Had we retained the 3 velocity
dispersion covariance parameters in our model (Sec. 4.7), our result
would have been 0.02 mag {\em brighter}.  Had we adopted the sample
Halo-2 rather than Halo-3, our result would have been $\sim$0.01 mag
{\em fainter}.


A final note in favor of our $M_V(RR)$ zero-point comes from
the derived kinematics (e.g., Sec. 3).  If the brighter Sandage (1993)
$M_V(RR)$ is used, the velocity dispersions for the local thick disk
and halo grow by 5-10\%.  The dispersions derived from our stat-$\pi$
solutions and presented in Table 6 are in good agreement with
estimates of the velocity ellipsoids of these components as measured
by other tracers (in fact, the $\sigma_U$ value for the halo is
already larger than many estimates).  Enlarging them to match the
brighter $M_V(RR)$ degrades this agreement.



\section{Consequences for Distances And Ages}

As discussed in Sec. 1, the $M_V(RR)$--[Fe/H] relation, in particular
its zero-point, is important in determining a number of quantities of
interest to both Galactic and extra-galactic astronomy.  In this
section, we present the implications of our stat-$\pi$--derived
$M_V(RR)$ zero-point.

In Sec. 5.4, we were unable to obtain a meaningful value for the slope
of the $M_V(RR)$--[Fe/H] relation.  However, our zero-point for the
halo RR Lyraes is quite well determined.  In the following discussion,
we adopt our zero-point along with a slope $\Delta
M_V(RR)/\Delta$[Fe/H] = 0.15 mag dex$^{-1}$, in agreement with HB
theory (Lee 1990), RGB theory (Fusi Pecci {\em et al.} 1990),
Baade-Wesselink observations (Jones {\em et al.} 1992), and some
analyses of the Sandage period shift effect (CSJ, Fernley 1993).
Specifically, we adopt the relation $M_V(RR)$ = 0.15[Fe/H] + 0.95 mag.

\subsection{Distance to the Galactic Center}

Walker \& Mack (1986) obtained the distance to the Galactic Center,
$R_0$, by finding the peak in the space density of RR Lyraes as a
function of distance along the line of sight through Baade's Window
(BW).  They recalibrated the photographic photometry of Blanco (1984)
to the Johnson $B$ band using several CCD standard fields, corrected
it for interstellar absorption, and converted it to the $V$ band using
a $\langle B \rangle_0 - \langle V \rangle_0 ~ vs.$ Period relation
obtained from NGC 6171.  They found $R_0 = 8.1 \pm 0.4$ if $M_V(RR)$ =
+0.60 mag.

We repeat their analysis here, comparing the results obtained using
both their preferred value of $M_V(RR)$ = +0.60 mag, and our preferred
value of $M_V(RR)$ = 0.15[Fe/H] + 0.95 mag.  We employ the newer
reddening estimates of $E(B-V)$ = 0.50 for BW (Walker \& Terndrup
1991) and $E(B-V)$ = 0.33 for NGC 6171 (Harris 1994).  We also employ
the Walker \& Terndrup (1991) $\Delta$$S$ metallicities, rather than
the photometric ones used by Walker \& Mack (1986).  Like those
authors, we found that a small shift (+0.04 mag) should be applied to
make the fitted line in the $\langle B \rangle_0 - \langle V \rangle_0
~ vs.$ Period plane coincide with the BW RR Lyraes listed in their
Table 7, presumably to correct for slight inconsistencies in the
adopted reddenings and/or metallicities of the cluster RR Lyraes
relative to those in BW.

Figure 8 shows the space density of stars (in arbitrary units) as a
function of distance through BW.  Using a variety of methods, we find
that the curve based on $M_V(RR)$ = 0.15[Fe/H] + 0.95 peaks at $R$ =
7.4 kpc (squares), while that based on $M_V(RR)$ = +0.60 peaks at $R$ =
8.1 kpc (circles).  
After applying the small geometric correction of 1.03 discussed by
Walker \& Mack (1986), we obtain $R_0 = 7.6 \pm 0.4$ and $R_0 = 8.3
\pm 0.5$, respectively.

\placefigure{fig8}

We note that the relation $M_V(RR)$ = 0.15[Fe/H] + 0.725 mag, based on
the Walker (1992) zero-point (see Table 7) produces results almost
identical with those of the $M_V(RR) = +0.60$ relation.  We also note
that the widths of the density curves produced using these three
$M_V(RR)$--[Fe/H] relations are nearly identical, after they are
corrected for the distance scale effects produced by the different
distance zero-points.  This supports the statement by Walker \&
Terndrup (1991) that the abundance dispersion in BW is too small to
meaningfully constrain the slope of the $M_V(RR)$--[Fe/H] relation
using this method.
 
The short distance to the Galactic Center based on our stat-$\pi$
zero-point is favored by the ``primary'' distance source, H$_2$O maser
proper motions ($R_0 = 7.2 \pm 1.3$ kpc), quoted by Reid (1993).  Reid
lists a number of other $R_0$ estimates, and from them derives a
``best value'' of $R_0 = 8.0 \pm 0.5$ kpc.  As noted by Carney {\em et
al.}  (1995), this value is in part determined using a ``bright'' RR
Lyrae calibration.  If we recompute the ``best value'' excluding the
optical RR Lyrae-dependent methods, we obtain $R_0 = 7.9 \pm 0.6$ kpc.
This value is further supported by Carney {\em et al.} (1995), who
recently found $R_0 = 7.8 \pm 0.4$ kpc from the $K$-band photometry of
58 RR Lyrae stars in BW, as calibrated using the
$M_K(RR)$--log($Period$) relation of CSJ.

\subsection{Ages of Globular Clusters}

The age of a globular cluster can be determined by comparing the
absolute magnitude of its main-sequence turnoff (MSTO) with the MSTOs
of theoretical isochrones.  The difference in apparent magnitude
between a cluster's RR Lyraes and its MSTO, together with an adopted
value for $M_V(RR)$, can be used to find the absolute magnitude of the
cluster's MSTO.

It is instructive to see the difference between the ages derived using
our $M_V(RR)$ zero-point and those using the brighter zero-point of
Walker (1992).  For both cases, we adopt a metallicity dependence of
0.15 mag dex$^{-1}$, so the relations are $M_V(RR) = 0.15$ [Fe/H] +
0.95, and $M_V(RR) = 0.15$ [Fe/H] + 0.725, respectively.  The
zero-point of the $M_V(RR)$--[Fe/H] relation primarily determines the
mean age of the globular cluster system, while the slope of the
relation determines the age distribution.  Thus our comparison focuses
on the mean ages of the halo globular cluster system under the two
$M_V(RR)$ zero-points.

Brian Chaboyer has kindly computed ages for the 39 ``older'' clusters
listed in Table 3 of Chaboyer, Demarque \& Sarajedini (1996b) using
both of these $M_V(RR)$--[Fe/H] relations.  The ages were derived
using the OPAL equation of state isochrones of Chaboyer \& Kim (1995).
These isochrones include the effects of diffusion and the latest
available equation of state (Rogers 1994), and employ modern helium
and $\alpha$-element abundances.

The weighted mean age of the 39 clusters is $14.8 \pm 0.2$ Gyr using
our zero-point, and $11.7 \pm 0.2$ Gyr using the Walker (1992)
zero-point (standard errors of the mean).  We account for the
uncertainties in the $M_V(RR)$ zero-points by computing the mean ages
from $M_V(RR)$--[Fe/H] relations based on zero-points of
$M_V(RR)\pm\sigma_{Mv}$ (again, thanks to B. Chaboyer).  For our
stat-$\pi$ $M_V(RR)$ zero-point, the mean age including formal errors
is $14.8 ^{+2.1}_{-1.8}$ Gyr,
compared with $11.7 ^{+1.4}_{-1.3}$ Gyr using the Walker (1992)
zero-point.  These values are $\sim$14\% smaller than the ages
computed without the improved treatments of diffusion and the equation
of state (Chaboyer {\em et al.} 1996b).  Clearly, the $M_V(RR)$
zero-point derived from the present stat-$\pi$ analysis supports an
older age for the globular cluster system.

The ages of the {\em oldest} globular clusters place a lower limit on
the age of the Universe.  Chaboyer {\em et al.}  (1996a) define a
group of 17 clusters which they suspect represents the oldest Galactic
globulars.  Using the method described above, their weighted mean age
is $16.5 ^{+2.1}_{-1.9}$ Gyr using our stat-$\pi$ $M_V(RR)$
zero-point, and $13.1 ^{+1.5}_{-1.3}$ Gyr using Walker's zero-point.
A word of caution is warranted here.  The median [Fe/H] of this group
is $\sim$0.2 dex lower than that of the field RR Lyraes used to
determine the $M_V(RR)$ zero-point, so if the $M_V(RR)$--[Fe/H] slope
is 0.30 (0.0) mag dex$^{-1}$, the mean derived ages are $\sim0.5$ Gyr
smaller (larger) than those quoted.

\subsection{Distance to the LMC and its Effect on $H_0$}

Walker (1992) lists the de-reddened $\langle V \rangle_I$ magnitudes
of the RR Lyrae stars in seven Large Magellanic Cloud (LMC) globular
clusters.  Using an LMC distance modulus of 18.50 $\pm$ 0.10, based on
LMC Cepheid observations and an abundance-corrected Galactic Cepheid
calibration, he obtained $M_V(RR) = +0.44$ at [Fe/H] = --1.9,
consistent with the ``brighter'' $M_V(RR)$ values shown in Fig. 7.
This distance modulus implies an LMC distance of $50 \pm 2$ kpc.

Using Walker's $\langle V \rangle_I$ magnitudes, and adopting $M_V(RR)
= 0.15$[Fe/H] + 0.95, based on our new stat-$\pi$ zero-point, we find
the LMC distance modulus to be 18.28 $\pm$ 0.13, equivalent to a
distance of 45 $\pm$ 3 kpc.

Given the complexities of deriving the Cepheid period-luminosity
relation (see Sec. 6, also Walker (1992) and references therein),
it seems worthwhile to explore the consequences of applying a
zero-point offset to the Cepheid period-luminosity relation to make it
match the shorter LMC distance based on our stat-$\pi$ results.

A particularly interesting consequence involves the measurement of the
Hubble constant, $H_0$.  Many extra-galactic distance indicators are
calibrated to, or are consistent with, an LMC distance modulus of
18.50 mag.  If the distances determined using these indicators are
recalibrated to agree with our smaller LMC distance modulus, the value
of $H_0$ derived from them increases by 10\%.  For example, the recent
Cepheid-based distance of $15.8 \pm 1.3$ Mpc to M100 yielded $H_0 = 83
\pm 16$ km~s$^{-1}$~Mpc$^{-1}$ (Ferrarese {\em et al.} 1996).  If we
reduce the M100 distance modulus by 0.22 mag to bring the Cepheid
period-luminosity relation into agreement with our LMC distance, we
obtain $d_{M100} = 14.3 \pm 1.2$ Mpc and $H_0 = 92 \pm 18$
km~s$^{-1}$~Mpc$^{-1}$.


Freedman {\em et al.} (1994; their Fig. 3) have already shown that the
expansion age of the Universe implied by $H_0 = 80 \pm 17$
km~s$^{-1}$~Mpc$^{-1}$ in the framework of the Einstein--de Sitter
cosmological model is in conflict with the observed ages of globular
clusters.  Our stat-$\pi$ $M_V(RR)$ zero-point indicates a larger
value of $H_0$ (shorter expansion time) and older globular clusters,
thus increasing the disagreement between these two important
observables.  The disagreement persists even at lower values of the
density parameter ($\Omega \approx 0.1$).  However, other recent
measurements of $H_0$ obtain lower values which are in better
agreement with our cluster ages.  For example, Branch {\em et al.}
(1996) obtained $H_0 = 57 \pm 4$ km~s$^{-1}$~Mpc$^{-1}$ from Type Ia
supernovae.


Additional possible routes to reconciling observations of $H_0$ and
globular cluster ages include accepting a non-zero cosmological
constant (Carroll \& Press 1992) and further refinements to stellar
evolution theory which result in younger cluster ages (e.g.,
Mazzitelli {\em et al.} 1995).

\section{Conclusions}

We have assembled high-quality data on 213 nearby RR Lyrae variables.
These data include new absolute proper motions from the Lick Northern
Proper Motion program (Klemola {\em et al.}  1993) and abundances and
radial velocities from Layden (1994). Based on an {\it a priori}
kinematic study, we defined three ways to separate the stars into
thick disk and halo sub-populations.  Statistical parallax solutions
for these sub-samples yielded the absolute magnitude and kinematics of
the RR Lyraes in the samples.  We note that our $M_V(RR)$ values
correspond to the absolute magnitude of typically-evolved RR Lyrae
stars, not to that of the Zero Age Horizontal Branch.

For the halo population, the solutions produced a well determined
absolute magnitude, $M_V(RR) = +0.71 \pm 0.12$ at
$\langle$[Fe/H]$\rangle$ = --1.61.  The derived kinematics, $\langle V
\rangle = -210 \pm 12~{\rm km~s}^{-1}$ and $(\sigma_U,\sigma_V,\sigma_W)
= (168 \pm 13,\ 102 \pm 8,\ 97 \pm 7)~{\rm km~s}^{-1}$,
are in good agreement with previous estimates of the halo RR Lyrae
kinematics, and with the kinematics of other stellar tracers of the
halo.

For the thick disk population, the results of the three definitions
scatter somewhat, and the uncertainties are larger.  Our best estimate
for the thick disk was $M_V(RR) = +0.79 \pm 0.30$ at
$\langle$[Fe/H]$\rangle$ = --0.76, after correction for the bias
mentioned below.  The derived kinematics, $\langle V \rangle = -48 \pm
9~{\rm km~s}^{-1}$ and $(\sigma_U,\sigma_V,\sigma_W) = ( 56 \pm 8,\ 51
\pm 8,\ 31 \pm 5)~{\rm km~s}^{-1}$,
again are in good agreement with previous estimates of the thick disk
kinematics based on RR Lyraes and on other tracers.

The large uncertainty in the disk solution prevented us from deriving a
meaningful slope for the $M_V(RR)$--[Fe/H] relation.  An attempt to
measure the slope by sub-dividing the halo sample into two metallicity
bins also failed to meaningfully determine the slope.

Monte Carlo tests using simulated data showed that our stat-$\pi$ code
accurately returns the true kinematic and $M_V$ values of the input
data.  They also revealed the possibility that the internal errors
returned by the HJBW stat-$\pi$ algorithm may be overestimates.  For
the halo, the true error of $M_V(RR)$ may be $\sim$0.08 mag rather than
0.12 mag.  The kinematic results for both the disk and halo may also be
more precise than the errors cited above.  However, it is outside the
scope of this paper to determine which set of error values is correct.  
So, for the present, we have conservatively adopted the larger values.

The simulations also enabled us to evaluate the effects of other
factors on the solutions.  All were negligible, except for a small
bias towards ``short'' distance scales that is observed when the $U$
and $V$ velocity dispersions are of comparable size.  We determined
corrections based on the simulations: --0.15 mag for the disk and
--0.04 mag for the halo solutions.  The former correction was applied
to the real-data solutions, while the latter was deemed too small to
be of practical value.

We discussed the effects of systematic errors.  The main systematic
uncertainty is in the adopted reddening scale.  Our scale (Burstein \&
Heiles 1982) makes our $M_V(RR)$ value $\sim$0.11 mag fainter than it
would have been using the Sturch (1966) reddenings, but SRM argue that
the former scale is preferred.  Our adopted photometric scale makes our
quoted $M_V(RR)$ value as bright as possible ($\sim$0.07 mag).
Several other minor systematics, including the halo bias correction
noted above, were also considered.  It is unlikely that these
systematic errors alone can account for the difference between our
$M_V(RR)$ value and the brighter ($\sim$0.2 mag) values found by some
authors (e.g., Buonanno {\em et al.} 1990; Sandage 1990b, 1993; Walker
1992).

Unlike the methods used by many authors, the stat-$\pi$ method is
independent of other distance calibrations.  It also depends on a
relatively simple and well-tested model (Galactic kinematics) in
comparison to the other methods, which employ model atmospheres,
stellar evolution theory, empirical color--temperature relations, RR
Lyrae mass determinations, etc.

We investigated the implications of our halo $M_V(RR)$ by using an
$M_V(RR)$--[Fe/H] relation based on an adopted slope of $\Delta
M_V/\Delta$[Fe/H] = 0.15 mag dex$^{-1}$ in combination with our halo
zero-point.  We found the distance to the Galactic Center to be $R_0 =
7.6 \pm 0.4$ kpc based on observations of the RR Lyraes in Baade's
Window, in good agreement with many other estimates of $R_0$.  Using
the ``brighter'' $M_V(RR)$ values, $R_0$ is 10\% larger.  Following
Chaboyer {\em et al.} (1996b), we found the mean age of the 17 oldest
Galactic globular clusters to be $16.5 _{-2.1}^{+1.9}$ Gyr, 3.4 Gyr
older than the mean age obtained using the brighter RR Lyrae
zero-point; this places an important lower limit on the age of the
Universe.  We found the distance modulus of the Large Magellanic Cloud
to be $18.28 \pm 0.13$ mag.  Any estimates of the Hubble Constant,
$H_0$, which are based on an LMC distance modulus of 18.50 mag (e.g.,
the Cepheid study of Ferrarese {\em et al.} 1996) increase by 10\% if
their distance scales are recalibrated to match our LMC distance.
This increase implies a younger age for the Universe, in conflict with
the older globular cluster ages derived from our $M_V(RR)$ value.  The
conflict is lessened or eliminated if the true value of $H_0$ is low,
if the cosmological constant is non-zero, or if further refinements to
stellar evolution theory result in younger cluster ages.

\acknowledgments

The authors acknowledge valuable discussions with Drs. Brian Chaboyer,
Jan Lub, Robert Schommer, and Alistair Walker.  The comments of an
anonymous referee improved several sections of the paper.  ACL
acknowledges financial support from the Natural Sciences and
Engineering Research Council of Canada, through research grants to
W.E. Harris and D.L. Welch, and from Cerro Tololo Inter-American
Observatory.  The Lick Northern Proper Motion program is supported by
National Science Foundation grant AST 92-18084.  SLH acknowledges the
support of NSF Young Investigator (NYI) grant number AST94-57455.  CJH
performed part of this research as an REU student at Michigan State
University.


\vskip 0.7in
\center{\bf Figure Captions}
\vskip 0.2in

\figcaption[Layden.fig1.ps]{The difference in proper motions measured
by the Lick NPM program and listed in the WMJ compilation (in the
sense NPM--WMJ) is plotted as a function of the proper motion error
given by WMJ, for (a) the right ascension component, and (b) the
declination component.  In each panel, the solid horizontal line
represents the mean proper motion difference for all stars plotted:
$-0\farcs11 \pm 0\farcs09 \ {\rm cent}^{-1}$ in $\Delta\mu_\alpha$
(105 stars) and $-0\farcs18 \pm 0\farcs08 \ {\rm cent}^{-1}$ in
$\Delta\mu_\delta$ (107 stars). \label{fig1}}

\figcaption[Layden.fig2.ps]{The running values of the nominal total
RMS error (labeled ``tot'') and of the actual RMS dispersion of the
proper motion differences (labeled ``$\Delta\mu$'') are plotted as a
function of the ``rank'' of the WMJ proper motion error (rank 1 =
smallest), for (a) the right ascension component, and (b) the
declination component.  In each panel, the short-dashed horizontal
line (``NPM'') represents the fixed contribution of the NPM proper
motion error $0\farcs5 \ {\rm cent}^{-1}$, and the long-dashed line
shows a constant RMS error ($0\farcs85 \ {\rm cent}^{-1}$ in
$\Delta\mu_\alpha$; $0\farcs80 \ {\rm cent}^{-1}$ in
$\Delta\mu_\delta$).  See Sec. 2.1 for further explanation. \label{fig2}}

\figcaption[Layden.fig3.ps]{The (a) radial, and (b) vertical
components of space velocity are plotted as a function of rotational
velocity for 213 RR Lyraes.  Squares indicate NPM proper motions,
while triangles indicate proper motions from WMJ.  Solid symbols have
[Fe/H] $> -1$, while open symbols have [Fe/H] $\leq -1$.\label{fig3} }

\figcaption[Layden.fig4.ps]{The rotational component of space
velocity is plotted as a function of abundance for 213 RR Lyraes.
Squares and triangles indicate NPM and WMJ proper motions,
respectively.  The sloping line separates disk and halo stars
according to Definition 1 (see Table 3), while the dash-dot line
indicates the separation according to Definition 2.  Filled symbols
indicate stars belonging to the disk, according to Definition 3, while
open symbols indicate stars belonging to the halo.  The vertical dotted
line separates the groups Halo-1P and Halo-1R. \label{fig4}}

\figcaption[Layden.fig5.ps]{Monte Carlo simulation results for 18 disk
data sets (filled squares) and 20 halo data sets (open squares),
listed respectively as solutions D2d and H2d in Table 5.  The difference
between the absolute magnitude computed using the stat-$\pi$ method
and that ``built into'' the data set is plotted as a function of the
velocity dispersion ratio $\sigma_U / \sigma_V$ (see Sec. 4.8).  The
solid line is the least-squares fit to the disk solutions, and the
dashed line marks the mean of the halo solutions.  The arrows mark the
velocity dispersion ratios of the real data (Disk-3 and Halo-3). \label{fig5}}

\figcaption[Layden.fig6.ps]{The stat-$\pi$ solutions for the real star
sub-samples listed in Table 6 are plotted in the abundance--magnitude
plane.  Circles mark the solutions for the Disk-1 and Halo-1
sub-samples, squares mark the Disk-2 and Halo-2 solutions, and
triangles mark the Disk-3 and Halo-3 solutions.  Filled symbols
indicate results after correction for the $\sigma_U / \sigma_V$ bias
discussed in Sec. 4.8, while open symbols (dotted error bars) indicate
pre-correction solutions.  The crosses mark the solutions for Halo-1P
and Halo-1R. \label{fig6}}

\figcaption[Layden.fig7.ps]{Our final stat-$\pi$ solutions for
$M_V(RR)$ are plotted against abundance (circles), together with the
results of other $M_V(RR)$ studies.  The symbol keys are given in the
right-most column of Table 7. \label{fig7}}

\figcaption[Layden.fig8.ps]{The space density of RR Lyraes (in
arbitrary units) as a function of distance along the line of sight
through Baade's Window.  Squares mark the distribution which employed
$M_V(RR)$ = 0.15[Fe/H] + 0.95, while circles mark the distribution
obtained using $M_V(RR)$ = +0.60 mag.  The dotted lines present an
alternate binning scheme for the two distributions. \label{fig8}}





\begin{deluxetable}{clcc}

\tablenum{1}

\tablewidth{35pc}
\tablecaption{RR Lyrae Photometry Correlations. \label{tbl-1}}
\tablehead{
\colhead{Source\tablenotemark{a}} & 
\colhead{$\langle V \rangle _{I,CD80}$} &
\colhead{Std Dev}   & \colhead{$N_{stars}$}
}
\startdata
1 & $0.146 + 0.983\cdot\langle V \rangle _{I,B77}$  
& 0.050 & 42 \nl
2 & $6.873 + 0.985\cdot2.5\cdot\langle V \rangle _{L} - 
0.034\cdot2.5\cdot\langle V-B \rangle _{L}$  &  0.054  & 51  \nl
3 & $-0.086  +  \langle V \rangle _{I,S}$  
& 0.064 &  13 \nl
\enddata
\tablenotetext{a}{Sources: 
1=Bookmyer {\em et al.} 1977;
2=Lub 1977;
3=Schmidt {\em et al.} 1991, 1995.
}
\end{deluxetable}



\begin{deluxetable}{rrrrrrrrrrrcc}
\tablenum{2}
\tablewidth{7.5in}
\tablecaption{Basic Data. \label{tbl-2}}

\tablehead{
\colhead{Star}         & \colhead{NPM1}    & \colhead{$l$} & 
\colhead{$b$} & \colhead{$\mu_{\alpha}$} & 
\colhead{$\mu_{\delta}$} & \colhead{$V_{rad}$} & 
\colhead{$\epsilon_V$} & \colhead{[Fe/H]} & \colhead{$\langle V \rangle_I$}  & 
\colhead{$A_V$} & \colhead{Ref}  & 
\colhead{D/H}  \\[.2ex]
\colhead{}   & \colhead{}   & \colhead{$^{\circ}$}   &
\colhead{$^{\circ}$}   & \colhead{$\arcsec$/cent}   &
\colhead{$\arcsec$/cent}   & \colhead{km/s}   &
\colhead{km/s}   & \colhead{dex}   & \colhead{mag}   &
\colhead{mag}   & \colhead{\tablenotemark{a}}   & \colhead{\tablenotemark{b}}
}

\startdata
SW And     &  +29.0017 & 115.72  &  --33.09  &    0.53 &  --2.49 & --21 &  2 & --0.38 &  9.68 & 0.14  & 1121 	 & 111 \nl
XX And     &  +38.0072 & 128.45  &  --23.64  &    3.53 &  --3.56 &    0 &  5 & --2.01 & 10.63 & 0.13  & 1111 	 & 000 \nl
XY And     &  +33.0068 & 131.22  &  --28.23  &    1.34 &  --0.28 & --64 & 53 & --0.92 & 13.63 & 0.15  & 1161	 & 000 \nl
ZZ And     &  +26.0036 & 122.42  &  --35.85  &    2.50 &  --1.80 & --13 & 53 & --1.58 & 13.01 & 0.12  & 1141	 & 000 \nl
CI And     &  +43.0105 & 134.93  &  --17.62  &  --0.15 &  --0.38 &   99 & 30 & --0.83 & 12.15 & 0.26  & 1141	 & 111 \nl
DR And     &  +33.0052 & 126.16  &  --28.57  &    3.00 &  --1.38 & --81 & 30 & --1.48 & 12.34 & 0.09  & 1141	 & 000 \nl
WY Ant     &  --~~~~~  & 266.93  &    22.08  &    2.56 &  --4.81 &  211 & 24 & --1.66 & 10.83 & 0.18  & 2111	 & 000 \nl
SW Aqr     & --00.1628 &  51.31  &  --31.47  &  --4.67 &  --6.08 & --42 &  8 & --1.24 & 11.14 & 0.22  & 1111 & 000 \nl
SX Aqr     &  +03.1345 &  57.90  &  --34.00  &  --3.87 &  --5.14 &--166 &  7 & --1.83 & 11.70 & 0.11  & 1111	 & 000 \nl
TZ Aqr     & --05.1879 &  53.25  &  --44.33  &    0.96 &  --1.77 & --35 & 12 & --1.24 & 12.01 & 0.10  & 1121	 & 000 \nl
YZ Aqr     & --11.1939 &  48.93  &  --49.76  &  --1.25 &    0.21 &--150 & 14 & --1.55 & 12.65 & 0.08  & 1161	 & 000 \nl
AA Aqr     & --10.2266 &  54.63  &  --53.83  &    2.11 &  --0.30 & --20 & 14 & --2.09 & 12.37 & 0.15  & 1161	 & 000 \nl
BN Aqr     & --07.2815 &  56.22  &  --50.73  &    0.96 &  --3.26 &--182 & 30 & --1.33 & 12.52 & 0.10  & 1161	 & 000 \nl
BO Aqr     & --12.2527 &  55.40  &  --58.83  &  --0.82 &  --1.21 & --24 & 13 & --1.80 & 12.11 & 0.07  & 1111 & 000 \nl
BR Aqr     & --09.2377 &  75.48  &  --65.24  &    0.50 &  --0.02 &   29 & 10 & --0.84 & 11.40 & 0.04  & 1121	 & 111 \nl
BT Aqr     & --05.1748 &  42.94  &  --30.60  &  --0.30 &  --1.06 & --52 & 11 & --0.29 & 12.34 & 0.12  & 1121 & 111 \nl
CP Aqr     & --01.1098 &  48.74  &  --31.34  &  --1.45 &  --1.73 &   29 & 21 & --0.90 & 11.71 & 0.11  & 1131	 & 111 \nl
DN Aqr     &  --~~~~~  &  35.76  &  --69.06  &    4.60 &  --1.60 &--214 &  8 & --1.63 & 11.18 & 0.02  & 2131 & 000 \nl
AA Aql     & --03.1592 &  43.08  &  --24.99  &  --1.80 &  --1.54 & --32 &  4 & --0.58 & 11.74 & 0.21  & 1121 & 111 \nl
V341 Aql   &  --~~~~~  &  45.62  &  --22.04  &    3.48 &  --2.40 & --81 &  4 & --1.37 & 10.81 & 0.31  & 2121	 & 000 \nl
X Ari      &  +10.0299 & 169.08  &  --39.84  &    5.97 &  --9.24 & --35 &  3 & --2.40 &  9.51 & 0.50  & 1111	 & 000 \nl
TZ Aur     &  +40.0228 & 176.79  &    20.92  &  --0.30 &  --1.35 &   46 &  6 & --0.80 & 11.86 & 0.18  & 1121	 & 111 \nl
RS Boo     &  +31.0704 &  50.84  &    67.35  &    0.22 &  --1.17 &  --9 &  2 & --0.32 & 10.36 & 0.00  & 1121	 & 111 \nl
ST Boo     &  +35.0720 &  57.39  &    55.22  &  --1.54 &  --1.67 &   13 &  4 & --1.86 & 10.98 & 0.04  & 1121	 & 000 \nl
SV Boo     &  +39.0683 &  68.75  &    65.51  &  --0.17 &  --2.28 &--131 & 22 & --1.55 & 13.12 & 0.00  & 1121	 & 000 \nl
SW Boo     &  +36.0643 &  62.52  &    67.74  &  --4.58 &    0.13 & --18 & 18 & --1.12 & 12.34 & 0.00  & 1111	 & 000 \nl
SZ Boo     &  +28.0840 &  41.93  &    65.50  &  --0.77 &  --0.85 & --38 & 21 & --1.68 & 12.60 & 0.01  & 1111	 & 000 \nl
TW Boo     &  +41.0778 &  71.06  &    62.85  &    0.34 &  --5.86 & --99 &  4 & --1.41 & 11.20 & 0.01  & 1121	 & 000 \nl
UU Boo     &  +35.0710 &  56.50  &    58.01  &    0.86 &  --3.38 &   10 & 28 & --1.92 & 12.22 & 0.01  & 1121	 & 000 \nl
UY Boo     &  +13.0991 & 354.24  &    68.81  &  --0.28 &  --4.51 &  144 &  3 & --2.49 & 10.80 & 0.00  & 1121	 & 000 \nl
RZ Cam     &  +67.0052 & 147.98  &    23.17  &    0.75 &  --0.57 &--266 & 26 & --1.01 & 12.73 & 0.18  & 1121	 & 000 \nl
RW Cnc     &  +29.0370 & 197.49  &    43.53  &    0.79 &  --4.17 & --85 &  7 & --1.52 & 11.85 & 0.03  & 1161	 & 000 \nl
SS Cnc     &  +23.0300 & 198.94  &    26.28  &  --0.77 &  --1.72 & --27 & 16 & --0.07 & 12.16 & 0.09  & 1121	 & 111 \nl
TT Cnc     &  +13.0361 & 212.10  &    28.38  &  --4.60 &  --4.00 &   49 &  5 & --1.58 & 11.24 & 0.12  & 1111	 & 000 \nl
AN Cnc     &  +15.0560 & 212.10  &    35.03  &  --0.51 &  --2.53 &   16 & 14 & --1.45 & 13.16 & 0.06  & 1141	 & 000 \nl
AQ Cnc     &  +12.0610 & 218.04  &    38.10  &  --1.15 &  --3.39 &  390 & 20 & --1.53 & 12.00 & 0.04  & 1141	 & 000 \nl
AS Cnc     &  +25.0305 & 197.89  &    31.23  &    2.79 &  --0.83 &  258 & 26 & --1.89 & 12.50 & 0.05  & 1141	 & 000 \nl
W CVn      &  +38.0637 &  71.82  &    70.96  &  --1.71 &  --0.89 &   18 & 21 & --1.21 & 10.52 & 0.00  & 1111	 & 001 \nl
Z CVn      &  +44.0884 & 124.00  &    73.35  &  --0.79 &  --3.97 &   14 & 10 & --1.98 & 11.93 & 0.00  & 1121	 & 000 \nl
RR CVn     &  +34.0596 & 154.05  &    81.09  &  --1.58 &  --3.17 &  --5 & 21 & --1.08 & 12.55 & 0.01  & 1111	 & 000 \nl
RU CVn     &  +31.0668 &  53.96  &    74.51  &  --2.97 &    0.22 & --27 & 21 & --1.37 & 11.96 & 0.00  & 1121	 & 000 \nl
RX CVn     &  +41.0711 &  87.08  &    71.53  &  --0.20 &    0.10 &--158 & 28 & --1.31 & 12.57 & 0.00  & 1121	 & 000 \nl
RZ CVn     &  +32.1089 &  61.59  &    77.15  &  --5.84 &  --0.36 & --12 &  7 & --1.92 & 11.42 & 0.00  & 1121	 & 000 \nl
SS CVn     &  +40.0608 &  83.85  &    72.63  &    1.24 &  --5.10 & --15 & 22 & --1.52 & 11.89 & 0.00  & 1121	 & 000 \nl
SV CVn     &  +37.0978 & 139.98  &    79.40  &    0.12 &  --2.55 &   29 & 28 & --2.20 & 12.59 & 0.00  & 1121	 & 000 \nl
SW CVn     &  +37.0987 & 134.84  &    79.80  &  --0.96 &  --1.98 & --18 & 21 & --1.53 & 12.74 & 0.00  & 1121	 & 000 \nl
UZ CVn     &  +40.0532 & 139.53  &    75.93  &    0.06 &  --2.32 & --38 & 26 & --2.34 & 12.02 & 0.00  & 1141	 & 000 \nl
AL CMi     &  +05.0339 & 214.43  &    15.35  &  --1.21 &  --0.52 &   46 & 21 & --0.85 & 12.01 & 0.09  & 1151	 & 111 \nl
RV Cap     & --15.2350 &  33.13  &  --35.54  &    2.42 & --10.57 &--106 &  7 & --1.72 & 10.92 & 0.11  & 1111	 & 000 \nl
IU Car     &  --~~~~~  & 269.59  &  --22.95  &  --1.30 &    0.60 &  328 & 18 & --1.85 & 11.91 & 0.39  & 2111	 & 000 \nl
V499 Cen   &  --~~~~~  & 315.10  &    18.12  &    2.10 &  --0.25 &  323 & 24 & --1.56 & 11.05 & 0.18  & 2111	 & 000 \nl
DX Cep     &  +83.0238 & 119.50  &    21.95  &    1.86 &    0.91 &  --6 & 30 & --1.83 & 12.67 & 0.35  & 1161	 & 000 \nl
RR Cet     &  +01.0121 & 143.54  &  --59.89  &  --0.05 &  --3.88 & --75 &  1 & --1.52 &  9.75 & 0.02  & 1111	 & 001 \nl
RU Cet     & --16.0129 & 134.27  &  --78.63  &    2.15 &  --1.00 &   57 &  8 & --1.60 & 11.60 & 0.01  & 1111	 & 000 \nl
RV Cet     & --11.0305 & 177.32  &  --64.40  &    2.58 &  --1.83 & --93 &  7 & --1.32 & 10.76 & 0.02  & 1111	 & 000 \nl
RX Cet     & --15.0060 & 102.48  &  --77.64  &  --2.89 &  --6.25 & --58 &  7 & --1.46 & 11.36 & 0.03  & 1111 & 000 \nl
RZ Cet     & --08.0299 & 178.21  &  --60.34  &    2.59 &    1.88 & --10 &  6 & --1.50 & 11.84 & 0.03  & 1121	 & 000 \nl
XZ Cet     & --16.0267 & 182.40  &  --70.75  &    4.14 &  --0.13 &  167 & 10 & --2.27 &  9.49 & 0.00  & 1161	 & 000 \nl
RY Col     &  --~~~~~  & 246.47  &  --35.05  &    3.60 &    1.80 &  482 & 15 & --1.11 & 10.86 & 0.01  & 2131	 & 000 \nl
S Com      &  +27.0979 & 213.16  &    85.84  &  --2.22 &  --2.01 & --55 &  4 & --2.00 & 11.55 & 0.02  & 1111	 & 000 \nl
V Com      &  +27.0931 & 208.67  &    80.85  &  --0.84 &    0.47 &   23 & 28 & --1.75 & 13.16 & 0.02  & 1121	 & 000 \nl
RY Com     &  +23.0619 & 342.56  &    85.06  &  --0.61 &  --1.78 & --31 &  8 & --1.65 & 12.30 & 0.04  & 1121	 & 000 \nl
TV CrB     &  +27.1359 &  41.33  &    56.51  &  --0.31 &  --1.01 &--157 & 48 & --2.33 & 11.61 & 0.07  & 1161	 & 000 \nl
W Crt      & --17.1378 & 276.00  &    40.47  &  --2.38 &  --1.48 &   65 & 13 & --0.50 & 11.51 & 0.09  & 1121	 & 111 \nl
X Crt      & --10.1333 & 278.87  &    49.49  &  --0.13 &  --3.55 &   79 &  4 & --1.75 & 11.48 & 0.00  & 1111	 & 000 \nl
XZ Cyg     &  --~~~~~  &  88.21  &    16.98  &    8.43 &  --2.15 &--119 & 14 & --1.52 &  9.72 & 0.33  & 2111	 & 000 \nl
DM Cyg     &  +31.0966 &  79.46  &  --12.41  &    1.34 &  --0.72 &   12 & 23 & --0.14 & 11.49 & 0.69  & 1121	 & 111 \nl
DX Del     &  +12.1761 &  58.47  &  --18.84  &    1.41 &  --1.17 & --45 &  3 & --0.56 &  9.92 & 0.32  & 1121	 & 111 \nl
RW Dra     &  +57.0740 &  87.39  &    40.60  &  --0.23 &  --0.82 &--108 & 22 & --1.40 & 11.57 & 0.00  & 1121	 & 001 \nl
SU Dra     &  +67.0274 & 133.44  &    48.27  &  --4.95 &  --7.26 &--167 &  1 & --1.74 &  9.78 & 0.00  & 1111	 & 000 \nl
SW Dra     &  +69.0237 & 127.27  &    47.33  &  --3.22 &  --0.93 & --30 &  1 & --1.24 & 10.49 & 0.04  & 1121	 & 001 \nl
WY Dra     &  +80.0305 & 113.08  &    25.14  &  --1.11 &    0.88 &  --6 & 30 & --1.66 & 12.67 & 0.24  & 1161	 & 000 \nl
XZ Dra     &  +64.0579 &  95.65  &    22.50  &    0.56 &  --0.71 & --30 &  2 & --0.87 & 10.18 & 0.22  & 1121	 & 111 \nl
AE Dra     &  +55.0697 &  84.35  &    25.41  &  --0.83\tablenotemark{c} &  --0.24\tablenotemark{c} &--243 & 30 & --1.54 & 12.65 & 0.14  & 1161	 & 000 \nl
BC Dra     &  +76.0341 & 107.95  &    28.48  &  --2.18 &    3.30 &--161 & 26 & --2.00 & 11.60 & 0.18  & 1161	 & 000 \nl
BD Dra     &  +77.0411 & 108.63  &    28.25  &  --2.88\tablenotemark{c} &    0.07\tablenotemark{c} &--253 & 30 & --1.74 & 12.69 & 0.10  & 1161	 & 000 \nl
BT Dra     &  +60.0453 &  99.41  &    51.21  &    0.13 &  --3.60 &--156 & 30 & --1.55 & 11.94 & 0.00  & 1161	 & 000 \nl
RX Eri     & --15.0672 & 214.26  &  --33.88  &  --1.56 &  --0.70 &   66 &  1 & --1.30 &  9.68 & 0.08  & 1111	 & 000 \nl
SV Eri     & --11.0414 & 194.26  &  --53.47  &    1.45 &  --3.95 & --12 &  9 & --2.04 &  9.95 & 0.19  & 1111	 & 000 \nl
XY Eri     & --13.0580 & 207.42  &  --41.69  &    1.13 &    0.68 &  221 & 11 & --2.08 & 13.02 & 0.08  & 1161	 & 000 \nl
BB Eri     & --19.0632 & 218.81  &  --34.36  &    3.45 &    0.91 &  235 & 11 & --1.51 & 11.46 & 0.03  & 1121	 & 000 \nl
BK Eri     & --01.0185 & 175.80  &  --51.70  &    3.24\tablenotemark{c} &  --1.98\tablenotemark{c} &  141 & 10 & --1.64 & 12.67 & 0.10  & 1161	 & 000 \nl
SS For     &  --~~~~~  & 216.42  &  --72.99  &    4.30 &  --7.25 &--112 &  1 & --1.35 & 10.10 & 0.00  & 2121 & 000 \nl
SW For     &  --~~~~~  & 243.27  &  --60.75  &    1.25 &  --0.10 &  174 & 18 & --1.95 & 12.34 & 0.00  & 2111	 & 000 \nl
RR Gem     &  --~~~~~  & 187.44  &    19.52  &  --0.35 &  --0.24 &   64 &  1 & --0.35 & 11.34 & 0.21  & 2121	 & 111 \nl
SZ Gem     &  +19.0314 & 201.85  &    22.08  &  --1.17 &  --3.40 &  307 & 11 & --1.81 & 11.66 & 0.08  & 1121	 & 000 \nl
TW Her     &  +30.0973 &  55.87  &    24.80  &  --0.32 &  --0.74 &    4 & 16 & --0.67 & 11.23 & 0.17  & 1121	 & 111 \nl
VZ Her     &  +36.0784 &  59.59  &    34.59  &  --2.44 &  --1.95 &--115 &  4 & --1.03 & 11.44 & 0.12  & 1121	 & 000 \nl
AF Her     &  +41.0868 &  65.15  &    41.64  &  --1.62 &  --0.72 &--268 &  9 & --1.94 & 12.82 & 0.00  & 1121	 & 000 \nl
AG Her     &  +40.0809 &  64.49  &    41.46  &  --2.15 &  --1.91 &--103 & 21 & --2.01 & 12.66 & 0.00  & 1121	 & 000 \nl
CW Her     &  +35.0792 &  58.01  &    38.99  &  --1.65 &    1.31 &--285 & 30 & --2.09 & 12.47 & 0.05  & 1141	 & 000 \nl
DL Her     &  +14.1435 &  36.28  &    26.60  &    1.13 &  --0.13 & --61 & 14 & --1.32 & 12.37 & 0.35  & 1141	 & 101 \nl
GY Her     &  +37.1381 &  60.70  &    41.71  &    0.29 &    1.13 &--157 & 53 & --1.92 & 12.57 & 0.01  & 1141	 & 000 \nl
V394 Her   &  +17.1675 &  40.01  &    27.39  &  --0.67 &    0.67 & --74 & 10 & --1.48 & 12.87 & 0.21  & 1161	 & 000 \nl
SV Hya     &  --~~~~~  & 297.08  &    36.59  &  --5.51 &    0.81 &  100 &  8 & --1.70 & 10.51 & 0.34  & 2111	 & 000 \nl
SZ Hya     & --09.0958 & 239.77  &    25.93  &  --0.66 &  --3.85 &  140 &  9 & --1.75 & 11.23 & 0.05  & 1121	 & 000 \nl
UU Hya     &  +04.0500 & 230.41  &    38.18  &  --2.51 &  --1.12 &  295 & 14 & --1.65 & 12.27 & 0.05  & 1121	 & 000 \nl
WZ Hya     & --12.1166 & 254.26  &    34.41  &    0.14 &  --1.49 &  304 &  8 & --1.30 & 10.82 & 0.26  & 1121	 & 000 \nl
XX Hya     & --15.1036 & 244.59  &    21.35  &    1.86 &  --2.92 &   32 & 20 & --1.33 & 11.89 & 0.15  & 1121	 & 000 \nl
DD Hya     &  +02.0597 & 219.85  &    19.30  &  --0.70 &  --1.19 &  153 & 25 & --1.00 & 12.18 & 0.06  & 1151	 & 011 \nl
DG Hya     & --05.0893 & 233.78  &    24.95  &  --1.08 &  --2.52 &  164 & 18 & --1.42 & 12.14 & 0.06  & 1121	 & 000 \nl
DH Hya     & --09.0936 & 238.03  &    22.96  &  --2.37 &  --0.67 &  355 &  8 & --1.55 & 12.13 & 0.05  & 1111	 & 000 \nl
ET Hya     & --08.0736 & 233.50  &    18.31  &  --1.66 &  --0.73 &  320 & 20 & --1.69 & 12.06 & 0.08  & 1151	 & 000 \nl
FY Hya     &  --~~~~~  & 318.76  &    31.37  &  --4.12 &  --0.04 &   82 & 24 & --2.33 & 12.46 & 0.15  & 2111 & 000 \nl
GL Hya     &  +02.0649 & 223.71  &    25.46  &  --1.34 &  --0.90 &  223 & 21 & --1.45 & 12.95 & 0.07  & 1161	 & 000 \nl
GO Hya     &  +06.0328 & 221.77  &    30.32  &  --0.23 &  --0.98 & --25 & 23 & --0.83 & 12.34 & 0.09  & 1141	 & 111 \nl
V Ind      &  --~~~~~  & 355.33  &  --43.12  &  --7.00 &  --9.00 &  202 &  3 & --1.50 &  9.92 & 0.05  & 2111 & 000 \nl
CQ Lac     &  +39.0934 &  93.95  &  --14.55  &    0.34 &  --0.15 &   20 & 30 & --2.04 & 12.43 & 0.45  & 1161	 & 000 \nl
RR Leo     &  +24.0416 & 208.42  &    53.10  &  --1.69 &  --1.26 &   88 &  1 & --1.57 & 10.68 & 0.09  & 1111	 & 001 \nl
RX Leo     &  +26.0471 & 209.43  &    70.51  &    0.38 &  --2.66 &--121 &  6 & --1.38 & 11.90 & 0.00  & 1111	 & 000 \nl
SS Leo     &  +00.0760 & 265.32  &    57.06  &  --2.38 &  --2.79 &  163 &  3 & --1.83 & 11.03 & 0.04  & 1111	 & 000 \nl
ST Leo     &  +10.0701 & 253.44  &    66.15  &  --0.56 &  --3.37 &  153 &  4 & --1.29 & 11.46 & 0.09  & 1121	 & 000 \nl
SU Leo     &  +08.0618 & 228.92  &    43.82  &    0.60 &  --0.86 & --81 & 30 & --1.41 & 13.55 & 0.02  & 1161	 & 000 \nl
SW Leo     & --02.1164 & 255.63  &    48.98  &  --0.74 &  --0.68 &   46 & 11 & --1.45 & 13.08 & 0.07  & 1161	 & 000 \nl
SZ Leo     &  +08.0731 & 243.93  &    57.83  &  --1.60 &  --2.55 &  185 &  4 & --1.86 & 12.35 & 0.04  & 1121	 & 000 \nl
TV Leo     & --05.1129 & 262.99  &    49.06  &    1.07 &    0.28 & --96 &  5 & --1.97 & 12.10 & 0.07  & 1111	 & 000 \nl
WW Leo     &  +07.0715 & 226.04  &    38.45  &  --0.03 &  --2.63 & --66 & 20 & --1.48 & 12.47 & 0.09  & 1121	 & 000 \nl
AA Leo     &  +10.0702 & 254.14  &    66.09  &  --0.26 &  --3.35 &   32 & 24 & --1.47 & 12.27 & 0.09  & 1121	 & 000 \nl
AE Leo     &  +17.0930 & 234.20  &    68.19  &    2.38 &  --1.25 & --53 & 10 & --1.71 & 12.52 & 0.00  & 1151	 & 000 \nl
AN Leo     &  +06.0502 & 253.35  &    60.72  &    0.29 &  --3.06 & --68 & 17 & --1.14 & 12.45 & 0.13  & 1141	 & 000 \nl
AX Leo     &  +12.0892 & 248.29  &    66.30  &  --1.85 &  --2.08 &  182 & 10 & --2.28 & 12.18 & 0.07  & 1141	 & 000 \nl
BT Leo     &  +18.0551 & 228.38  &    65.59  &  --0.90 &  --0.34 &  119 & 14 & --0.81 & 13.11 & 0.00  & 1141	 & 111 \nl
V LMi      &  +29.0467 & 201.30  &    57.84  &    2.16 &  --3.02 &--110 &  7 & --1.15 & 11.71 & 0.02  & 1121	 & 000 \nl
X LMi      &  +39.0396 & 182.53  &    53.70  &    1.63 &  --2.01 & --82 & 18 & --1.68 & 12.31 & 0.00  & 1121	 & 000 \nl
U Lep      & --21.0669 & 221.10  &  --34.37  &    4.33 &  --5.86 &  128 & 11 & --1.93 & 10.60 & 0.02  & 1111	 & 000 \nl
RY Lib     & --21.1628 & 330.87  &    36.08  &  --1.68 &  --0.45 &   33 & 11 & --1.48 & 13.15 & 0.25  & 1161	 & 000 \nl
TV Lib     & --08.1548 & 353.16  &    39.67  &    0.05 &    1.04 & --61 & 10 & --0.27 & 11.94 & 0.25  & 1121	 & 111 \nl
VY Lib     & --15.2271 & 353.86  &    28.84  &  --0.22 &  --6.46 &  142 & 10 & --1.32 & 11.72 & 0.45  & 1111	 & 000 \nl
TW Lyn     &  +43.0251 & 176.15  &    27.54  &    0.69 &    0.86 & --40 & 26 & --1.23 & 11.90 & 0.14  & 1141	 & 101 \nl
Y Lyr      &  +43.0988 &  72.67  &    20.87  &  --0.08 &  --0.66 & --65 & 23 & --1.03 & 13.28 & 0.18  & 1121	 & 101 \nl
RR Lyr     &  --~~~~~  &  74.96  &    12.30  & --10.95 & --19.42 & --63 &  8 & --1.37 &  7.74 & 0.13  & 2111 & 000 \nl
RZ Lyr     &  +32.1588 &  62.11  &    15.82  &    1.02 &    1.99 &--233 & 23 & --2.13 & 11.51 & 0.32  & 1111	 & 000 \nl
CN Lyr     &  +28.1070 &  58.01  &    14.70  &  --0.12 &  --1.61 &   67 & 30 & --0.26 & 11.49 & 0.62  & 1161	 & 111 \nl
CX Lyr     &  +28.1083 &  58.99  &    12.72  &  --0.49 &  --1.30 &--203 & 30 & --1.79 & 12.83 & 0.82  & 1141	 & 000 \nl
IO Lyr     &  +32.1543 &  60.59  &    19.98  &  --1.22 &    2.19 &--157 & 30 & --1.52 & 11.86 & 0.15  & 1161	 & 000 \nl
UV Oct     &  --~~~~~  & 308.40  &  --23.55  &  --6.93 & --12.30 &  126 & 12 & --1.61 &  9.42 & 0.21  & 2111 & 000 \nl
ST Oph     &  --~~~~~  &  22.83  &    16.64  &  --0.09 &  --0.08 &   12 &  7 & --1.30 & 12.05 & 0.60  & 2111 & 101 \nl
V413 Oph   & --10.1983 &   4.39  &    25.97  &  --1.10 &  --1.62 & --39 & 30 & --1.00 & 12.08 & 0.63  & 1161 & 111 \nl
V445 Oph   &  --~~~~~  &   7.91  &    28.45  &    0.47 &    1.24 & --22 &  5 & --0.23 & 10.99 & 0.59  & 2121	 & 111 \nl
V452 Oph   &  +11.1284 &  32.52  &    25.72  &  --0.36 &  --0.10 &--375 & 30 & --1.72 & 12.18 & 0.41  & 1121	 & 000 \nl
V964 Ori   & --02.0731 & 202.50  &  --23.91  &    0.70 &  --1.21 &  178 & 11 & --1.89 & 12.95 & 0.25  & 1151	 & 000 \nl
TY Pav     &  --~~~~~  & 330.55  &  --17.10  &  --2.10 &  --2.20 &  245 &  9 & --2.31 & 12.58 & 0.26  & 2111 & 000 \nl
DN Pav     &  --~~~~~  & 332.82  &  --30.80  &  --0.90 & --3.00 & --69 & 12 & --1.54 & 12.42 & 0.14  & 2111 & 000 \nl
VV Peg     &  +18.1195 &  78.42  &  --30.42  &  --0.06 &  --1.22 &   13 &  8 & --1.88 & 11.79 & 0.13  & 1111	 & 000 \nl
AV Peg     &  +22.1796 &  77.44  &  --24.05  &    1.35 &  --1.34 & --58 &  1 & --0.14 & 10.44 & 0.14  & 1121	 & 111 \nl
BH Peg     &  +15.1616 &  85.62  &  --38.36  &  --2.01 &  --6.71 &--278 &  2 & --1.38 & 10.44 & 0.20  & 1111	 & 000 \nl
CG Peg     &  +24.0966 &  77.18  &  --20.75  &  --0.11 &  --0.49 &  --4 &  4 & --0.48 & 11.11 & 0.20  & 1121	 & 111 \nl
DZ Peg     &  +15.1715 &  93.09  &  --41.46  &    1.70 &  --2.49 &--294 & 11 & --1.52 & 12.00 & 0.05  & 1161	 & 000 \nl
GV Peg     &  +26.1124 & 109.07  &  --34.83  &    0.69 &  --3.07 &--335 & 30 & --1.99 & 13.36 & 0.10  & 1161	 & 000 \nl
AR Per     &  --~~~~~  & 154.93  &   --2.27  &  --0.15 &    0.50 &    5 &  1 & --0.43 & 10.43 & 1.08  & 2122	 & 111 \nl
U Pic      &  --~~~~~  & 257.67  &  --39.61  &  --0.10 &  --1.70 &   30 & 12 & --0.73 & 11.32 & 0.00  & 2131 & 111 \nl
RY Psc     & --02.0022 & 100.68  &  --62.89  &    3.99 &  --0.77 &  --1 &  9 & --1.39 & 12.28 & 0.08  & 1121	 & 000 \nl
BB Pup     & --19.0790 & 241.28  &    10.27  &  --1.58 &    1.05 &   98 &  9 & --0.57 & 12.17 & 0.45  & 1111	 & 111 \nl
V440 Sgr   &  --~~~~~  &  15.31  &  --19.20  &  --0.20 & --5.00 & --62 &  1 & --1.47 & 10.24 & 0.36  & 2121 & 000 \nl
RU Scl     &  --~~~~~  &  41.53  &  --78.86  &    5.63 &  --2.04 &   38 &  8 & --1.25 & 10.21 & 0.03  & 2111	 & 000 \nl
VY Ser     &  +01.1004 &   6.16  &    44.09  &  --9.84 &  --0.51 &--145 &  1 & --1.82 & 10.13 & 0.06  & 1111	 & 000 \nl
AN Ser     &  +13.1114 &  23.80  &    45.24  &  --0.27 &  --1.11 & --47 &  4 & --0.04 & 10.97 & 0.09  & 1121	 & 111 \nl
AR Ser     &  +02.1454 &   7.89  &    44.25  &  --3.66 &    0.60 &  132 &  4 & --1.78 & 11.85 & 0.07  & 1131	 & 000 \nl
AT Ser     &  +08.1225 &  18.03  &    42.45  &  --0.14 &  --1.48 & --58 & 11 & --2.05 & 11.45 & 0.08  & 1111	 & 000 \nl
AV Ser     &  +00.1096 &  11.28  &    36.83  &    0.80 &    0.65 & --45 & 13 & --1.20 & 11.40 & 0.26  & 1111	 & 101 \nl
AW Ser     &  +15.1229 &  28.67  &    43.35  &  --0.97 &  --1.63 &--126 & 15 & --1.67 & 12.79 & 0.04  & 1141	 & 000 \nl
BH Ser     &  +19.0930 &  27.56  &    56.27  &  --0.78 &  --1.51 &--113 & 11 & --1.59 & 12.85 & 0.08  & 1161	 & 000 \nl
CS Ser     &  +03.1110 &   7.22  &    45.43  &    2.37 &  --2.79 &    2 & 14 & --1.57 & 12.39 & 0.08  & 1141	 & 000 \nl
DF Ser     &  +18.0911 &  26.27  &    55.92  &    0.46 &  --0.55 & --10 & 14 & --0.74 & 12.69 & 0.07  & 1141	 & 111 \nl
RV Sex     & --08.0980 & 258.12  &    43.38  &  --0.98 &  --0.85 &  120 & 20 & --1.10 & 12.30 & 0.02  & 1161	 & 000 \nl
SS Tau     &  +05.0288 & 180.09  &  --38.53  &    0.77 &    0.28 & --11 & 10 & --0.28 & 12.50 & 0.49  & 1121	 & 111 \nl
U Tri      &  +33.0093 & 137.89  &  --27.24  &    0.89 &  --1.30 &    6 & 23 & --0.79 & 12.60 & 0.10  & 1121	 & 111 \nl
W Tuc      &  --~~~~~  & 301.66  &  --53.72  &    0.30 &    0.20 &   63 &  3 & --1.64 & 11.43 & 0.00  & 2111	 & 000 \nl
YY Tuc     &  --~~~~~  & 325.32  &  --54.21  &    0.14 &  --0.34 &   56 &  9 & --1.82 & 11.98 & 0.00  & 2111	 & 000 \nl
RV UMa     &  +54.0419 & 109.75  &    62.06  &  --2.76 &  --4.69 &--183 &  9 & --1.19 & 10.78 & 0.01  & 1111	 & 000 \nl
TU UMa     &  +30.0521 & 198.80  &    71.87  &  --7.64 &  --4.97 &   88 &  1 & --1.44 &  9.81 & 0.00  & 1111	 & 000 \nl
AB UMa     &  +48.0617 & 141.04  &    67.86  &  --1.30 &  --2.00 & --56 & 26 & --0.72 & 10.80 & 0.00  & 1141	 & 111 \nl
ST Vir     & --00.1211 & 346.37  &    53.65  &  --0.06 &  --2.65 & --22 & 13 & --0.88 & 11.52 & 0.07  & 1121 & 111 \nl
UU Vir     &  --~~~~~  & 280.73  &    60.52  &  --2.96 &  --0.45 & --8 &  1 & --0.82 & 10.56 & 0.01  & 2121 & 111 \nl
UV Vir     &  +00.0808 & 286.55  &    62.28  &  --2.63 &  --1.80 &   99 & 11 & --1.19 & 11.83 & 0.02  & 1121	 & 000 \nl
WY Vir     & --06.1416 & 321.78  &    54.28  &  --1.74 &  --1.17 &  181 & 10 & --2.84 & 13.38 & 0.03  & 1161	 & 000 \nl
AD Vir     & --07.2115 & 333.18  &    51.21  &  --1.97 &  --0.60 &  134 & 14 & --1.15 & 13.04 & 0.04  & 1151	 & 000 \nl
AE Vir     &  +04.0956 & 351.61  &    57.27  &  --0.03 &  --1.77 &  208 & 10 & --1.16 & 13.26 & 0.02  & 1141	 & 000 \nl
AF Vir     &  +06.0757 & 355.48  &    59.16  &  --6.16 &    0.05 & --35 & 14 & --1.46 & 11.52 & 0.01  & 1121	 & 000 \nl
AM Vir     & --16.1465 & 313.94  &    45.52  &  --0.16 &  --5.15 &   99 & 24 & --1.45 & 11.49 & 0.14  & 1121	 & 000 \nl
AS Vir     & --09.1409 & 303.47  &    52.61  &    1.14 &  --3.69 &   70 & 23 & --1.49 & 11.90 & 0.08  & 1121	 & 000 \nl
AT Vir     & --05.1349 & 304.66  &    57.40  &  --5.42 &  --1.76 &  346 &  8 & --1.91 & 11.27 & 0.04  & 1121	 & 000 \nl
AV Vir     &  +09.0882 & 325.01  &    70.82  &    0.64 &  --3.38 &  152 &  4 & --1.32 & 11.78 & 0.00  & 1121	 & 000 \nl
BQ Vir     & --02.1373 & 295.38  &    60.23  &  --0.27 &  --1.39 &  129 &  9 & --1.32 & 12.48 & 0.03  & 1161	 & 000 \nl
DO Vir     & --05.1546 & 345.60  &    48.45  &  --2.72 &    0.69 &   24 & 36 & --0.80 & 14.14 & 0.10  & 1151	 & 000 \nl
FU Vir     &  +13.0858 & 290.13  &    75.56  &    1.33 &  --0.77 & --90 &  8 & --1.17 & 12.63 & 0.07  & 1161	 & 000 \nl
FK Vul     &  +22.1711 &  67.60  &  --13.92  &    0.05 &  --1.51 & --76 & 30 & --0.95 & 12.87 & 0.35  & 1161	 & 111 \nl
AT And     &  --~~~~~  & 109.76  &  --18.09  &  --0.20 &    4.60 &--241 & 11 & --0.97 & 10.66 & 0.38  & 2221	 & 000 \nl
S Ara      &  --~~~~~  & 343.38  &  --12.45  &  --2.34 &  --1.53 &  172 & 13 & --1.43 & 10.67 & 0.36  & 2231 & 000 \nl
RU Boo     &  +23.0728 &  30.94  &    63.87  &  --1.35 &  --0.32 & --60 & 35 & --1.50 & 13.60 & 0.04  & 1321	 & 000 \nl
BI Cen     &  --~~~~~  & 294.66  &     2.44  &  --0.76 &    0.15 &  210 & 30 & --0.83 & 11.86 & 0.59  & 2262	 & 000 \nl
UU Cet     & --17.0006 &  73.26  &  --75.09  &    2.68 &  --0.65 &--114 &  3 & --1.32 & 11.95 & 0.01  & 1221	 & 000 \nl
Z Com      &  +18.0747 & 328.12  &    80.58  &  --0.77 &  --1.85 & --50 & 35 & --1.50 & 13.73 & 0.03  & 1321	 & 000 \nl
ST Com     &  --~~~~~  & 347.87  &    81.25  &  --3.61 &  --3.57 & --68 &  7 & --1.26 & 11.38 & 0.04  & 2221 & 000 \nl
SW Cru     &  --~~~~~  & 296.49  &     1.91  &    1.07 &    0.19 & --23 & 30 & --0.54 & 12.33 & 1.18  & 2262	 & 111 \nl
UY Cyg     &  --~~~~~  &  74.54  &   --9.63  &    0.13 &  --0.76 &  --2 &  6 & --1.03 & 11.05 & 0.22  & 2222	 & 101 \nl
SW Her     &  +21.1016 &  41.68  &    34.00  &  --1.11 &    0.04 &--130 & 35 & --1.50 & 14.14 & 0.21  & 1321	 & 000 \nl
VX Her     &  +18.0988 &  35.22  &    39.08  &  --4.70 &    1.66 &--377 &  3 & --1.52 & 10.62 & 0.18  & 1211	 & 000 \nl
AR Her     &  +47.1123 &  74.10  &    48.20  &  --6.53 &    1.24 &--349 &  8 & --1.40 & 11.18 & 0.04  & 1211	 & 000 \nl
RV Leo     &  --~~~~~  & 232.37  &    51.14  &  --0.50 &  --1.30 &    0 & 35 & --1.50 & 13.85 & 0.08  & 2321	 & 000 \nl
TT Lyn     &  +44.0496 & 176.07  &    41.65  &  --8.41 &  --4.01 & --67 &  1 & --1.76 &  9.87 & 0.03  & 1261	 & 000 \nl
EZ Lyr     &  --~~~~~  &  65.52  &    16.25  &  --1.32 &  --0.20 & --60 & 23 & --1.56 & 11.60 & 0.21  & 2261 & 001 \nl
AO Peg     &  +18.1149 &  69.90  &  --22.60  &  --0.31 &  --2.94 &  115 & 35 & --0.92 & 12.83 & 0.19  & 1261	 & 000 \nl
TU Per     &  --~~~~~  & 142.78  &   --4.29  &    1.51 &  --0.61 &--377 & 11 & --1.50 & 12.53 & 1.38  & 2324	 & 000 \nl
RV Phe     &  --~~~~~  & 336.01  &  --64.00  &    4.15 &  --1.85 & --99 &  2 & --1.60 & 11.75 & 0.06  & 2211 & 000 \nl
XX Pup     &  --~~~~~  & 236.65  &     8.72  &  --3.13 &  --0.14 &  386 &  7 & --1.50 & 11.20 & 0.38  & 2324 & 000 \nl
V675 Sgr   &  --~~~~~  & 358.26  &   --7.83  &    0.00 &    1.20 &--105 & 30 & --2.01 & 10.36 & 0.19  & 2212	 & 000 \nl
V1640 Sgr  &  --~~~~~  &   0.47  &  --13.64  &  --0.40 &    0.90 &   41 & 10 & --0.54 & 12.68 & 0.31  & 2251	 & 111 \nl
V494 Sco   &  --~~~~~  & 357.23  &   --0.49  &  --0.34 &  --0.62 &   26 & 30 & --1.01 & 11.27 & 1.16  & 2233 & 101 \nl
AF Vel     &  --~~~~~  & 284.16  &     8.60  &    5.90 &  --1.75 &  236 & 16 & --1.64 & 11.34 & 0.61  & 2214	 & 000 \nl
BB Vir     &  +06.0723 & 340.31  &    64.84  &  --3.71 &  --1.02 & --38 & 13 & --1.61 & 11.07 & 0.00  & 1221	 & 000 \nl
BC Vir     &  +06.0660 & 323.42  &    67.52  &    1.55 &  --2.81 &    4 & 13 & --1.50 & 12.21 & 0.00  & 1361	 & 000 \nl
BN Vul     &  --~~~~~  &  58.63  &     3.41  &  --4.85 &  --3.80 &--267 &  4 & --1.52 & 11.08 & 1.36  & 2262 &  000 
\enddata
\tablenotetext{a}{References:  First digit indicates source of proper motions
(1=NPM; 2=WMJ).  Second digit indicates source of abundance 
(1=L94, Table 9; 2=$\Delta$$S$ from L94, Table 2 and Eqn. 6; 3=[Fe/H]=--1.5).  
Third digit indicates source of photometry (1=CD80; 2=Bookmyer 
{\em et al} 1977; 3=Lub 1979; 4=Schmidt {\em et al} 1991, 1995; 
5=Layden 1996; 6=L94).  Fourth digit indicates source of $A_V$
(1=Burstein \& Heiles 1982; 2=Blanco 1992; 3=FitzGerald 1968,1987; 
4=interpolation).}
\tablenotetext{b}{Disk/Halo separation:  First digit indicates status by 
Definition 1 (1=disk, 0=halo).  Second digit indicates status by Definition 
2.  Third digit indicates status by Definition 3.}
\tablenotetext{c}{Proper motion changed from value in NPM1; one 
discordant proper motion value removed.}
\end{deluxetable}





\begin{deluxetable}{cl}
\tablenum{3}
\tablewidth{37pc}
\tablecaption{Disk/Halo Definitions. \label{tbl-3}}
\tablehead{
\colhead{Definition}         & \colhead{Description}  
}
\startdata
Disk-1 & All stars lying above/rightward of $V_{\theta} = -400$[Fe/H]$ - 300$ (see Fig. 3).\tablenotemark{a} \nl
Halo-1 & All stars lying below/leftward of this line.\tablenotemark{a} \nl
      &  \nl
Disk-2 & All stars having [Fe/H] $\geq -1.0$ and $V_{\theta} > 80$ km~s$^{-1}$.\tablenotemark{b} \nl
Halo-2 & All stars excluded from Disk-2.\tablenotemark{b} \nl
      &  \nl
Disk-3 & All stars in Disk-1 $plus$ all stars having $|V_{\pi}| < 100$ 
km~s$^{-1}$, $V_{\theta} > 80$ km~s$^{-1}$, \nl
      & ~~~~$|V_{z}| < 60$ km~s$^{-1}$, $|Z|$ $< 1.0$ kpc, $and$ 
[Fe/H] $> -1.6$. \nl
Halo-3 & All stars excluded from Disk-3. \nl
      &  \nl
Halo-1R & All stars in Halo-1 with [Fe/H] $> -1.55$. \nl
Halo-1P & All stars in Halo-1 with [Fe/H] $\leq -1.55$. \nl
\enddata
\tablenotetext{a}{Two stars, AO Peg and FU Vir, lie above/rightward 
of the line defining Disk-1, yet their extreme $V_{\pi}$ and $V_z$
velocities clearly indicate that they belong to the halo.  They were
removed from Disk-1 and included in Halo-1.}
\tablenotetext{b}{AO Peg fits the definition for Disk-2, yet clearly 
belongs to the halo.  It was removed from Disk-2 and included in Halo-2.}
\end{deluxetable}





\begin{deluxetable}{ccccccccc}
\tablenum{4}
\tablewidth{39pc}
\tablecaption{Simulated Data Sets. \label{tbl-4}}
\tablehead{
\colhead{Data} & \colhead{Population} & 
\colhead{Space} & \colhead{$N_{stars}$} & 
\colhead{$N_{trials}$}  & \colhead{Solution} &
\colhead{$\sigma_{k}$}  & \colhead{VE-covar}  &
\colhead{$N_{conv}$}  \\[.2ex]
\colhead{Set} & \colhead{Simulated} & 
\colhead{Distribution} & \colhead{} & 
\colhead{}  & \colhead{Set} &
\colhead{}  & \colhead{included}  &
\colhead{} 
}
\startdata

H1  & halo\tablenotemark{a}  & random    & 165  & 5  & H1.0  & 0.0  & yes & 5  \nl
    &                        &           &      &    & H1    & 0.1  & yes & 5  \nl
H2  & halo\tablenotemark{a}  & real      & 169  & 5  & H2.0  & 0.0  & yes & 5  \nl
    &                        &           &      &    & H2    & 0.1  & yes & 5  \nl
    &                        &           &      &(+15)& H2d  & 0.1  & no  & 20 \nl
    &                        &           &      &    &       &      &     &    \nl
D1  & disk\tablenotemark{b}  & random    & 50   & 20 & D1    & 0.1  & yes & 19 \nl
D2  & disk\tablenotemark{b}  &   real    & 45   & 20 & D2    & 0.1  & yes & 18 \nl
    &                        &           &      &    & D2d   & 0.1  & no  & 18 \nl

\enddata
\tablenotetext{a}{Halo $V_{(\pi,\theta,z)} = (0, 20, 0)$ km s$^{-1}$; 
$\sigma_{(\pi,\theta,z)} = (160, 100, 90)$ km s$^{-1}$.}
\tablenotetext{b}{Disk $V_{(\pi,\theta,z)} = (0, 200, 0)$ km s$^{-1}$; 
$\sigma_{(\pi,\theta,z)} = (50, 50, 30)$ km s$^{-1}$.}
\end{deluxetable}





\begin{deluxetable}{crrrrrrrrrrrr}
\tablenum{5}
\tablewidth{43pc}
\tablecaption{Monte Carlo Simulation Results. \label{tbl-5}}
\tablehead{
\colhead{Solution}  & \colhead{} & \colhead{$V$} & \colhead{$\sigma_{U}$} & 
\colhead{$\sigma_{V}$} & \colhead{$\sigma_{W}$} & \colhead{$M_V$} &
\colhead{} &
\colhead{$\Delta V$} & \colhead{$\Delta \sigma_{U}$} & 
\colhead{$\Delta \sigma_{V}$} & \colhead{$\Delta \sigma_{W}$} & 
\colhead{$\Delta M_V$}
}
\startdata

H1.0  &mean&--203 & 158 & 100 &  89 &+0.82 &mean&  +3 & --2 & --1 & --3 & +0.05 \nl
 &$\langle \sigma_i \rangle$ &   11 &  11 &   8 &   7 & 0.12 & SD &   5 &   2 &   7 &   3 &  0.07 \nl
 & & & & & & & & & & & & \nl
H1    &mean&--204 & 158 &  99 &  89 &+0.79 &mean&  +2 & --2 & --3 & --3 & +0.02 \nl
 &$\langle \sigma_i \rangle$ &   12 &  11 &   8 &   7 & 0.12 & SD &   5 &   2 &   6 &   3 &  0.07 \nl
 & & & & & & & & & & & & \nl
H2.0  &mean&--198 & 155 &  96 &  86 &+0.84 &mean&  +7 & --7 & --3 & --2 & +0.07 \nl
 &$\langle \sigma_i \rangle$ &   11 &  12 &   7 &   6 & 0.12 & SD &   2 &   5 &   3 &   4 &  0.08 \nl
 & & & & & & & & & & & & \nl
H2    &mean&--199 & 155 &  94 &  86 &+0.82 &mean&  +6 & --6 & --4 & --2 & +0.05 \nl
 &$\langle \sigma_i \rangle$ &   12 &  12 &   7 &   6 & 0.12 & SD &   2 &   5 &   3 &   3 &  0.08 \nl
 & & & & & & & & & & & & \nl
H2d   &mean&--205 & 155 &  93 &  87 &+0.81 &mean&  +4 & --4 & --4 & --2 & +0.04 \nl
 &$\langle \sigma_i \rangle$ &   12 &  12 &   7 &   6 & 0.12 & SD &   5 &   6 &   3 &   3 &  0.08 \nl
 & & & & & & & & & & & & \nl
 & & & & & & & & & & & & \nl
D1    &mean& --32 &  51 &  48 &  29 &+0.88 &mean&   0 & --2 & --1 & --2 &--0.02 \nl
 &$\langle \sigma_i \rangle$ &    8 &   8 &   8 &   6 & 0.34 & SD &   5 &   7 &   7 &   7 &  0.44 \nl
 & & & & & & & & & & & & \nl
D2    &mean& --33 &  51 &  48 &  27 &+1.04 &mean&  +1 & --4 & --4 & --2 & +0.13 \nl
 &$\langle \sigma_i \rangle$ &    9 &   9 &   8 &   6 & 0.35 & SD &   5 &   6 &   6 &   5 &  0.29 \nl
 & & & & & & & & & & & & \nl
D2d   &mean& --32 &  51 &  48 &  26 &+1.04 &mean&  +1 & --4 & --4 & --3 & +0.13 \nl
 &$\langle \sigma_i \rangle$ &    9 &   8 &   8 &   6 & 0.34 & SD &   5 &   5 &   7 &   4 &  0.29 \nl
\enddata
\end{deluxetable}





\begin{deluxetable}{lrcrrrrrrrr}
\tablenum{6}
\tablewidth{42pc}
\tablecaption{Statistical Parallax Solutions. \label{tbl-4}}
\tablehead{
\colhead{Sample} &  \colhead{$N_{stars}$}  & 
\colhead{$\langle$[Fe/H]$\rangle$} &
\colhead{$\langle U \rangle$} &  \colhead{$\langle V \rangle$} & 
\colhead{$\langle W \rangle$} &  \colhead{$\sigma_{U}$} & 
\colhead{$\sigma_{V}$} &  \colhead{$\sigma_{W}$} & \colhead{$M_V$} &
\colhead{$M_{V,corr}$} \\[.3ex]
\colhead{} &  \colhead{}  & \colhead{dex} &
\colhead{km/s} & \colhead{km/s} & \colhead{km/s} & 
\colhead{km/s} & \colhead{km/s} & \colhead{km/s} & 
\colhead{mag} & \colhead{mag} 
}
\startdata
Halo-1    & 169 &--1.60 &  +9 &--205 &--11 & 165 & 102 & 95 & +0.71 & +0.71 \nl
 & & &              $\pm$~ 13 &   12 &   8 &  12 &   7 &  7 &  0.12 &  0.12 \nl
      & & & & & & & & & & \nl
Halo-2    & 175 &--1.58 &  +8 &--196 &--11 & 161 & 108 & 93 & +0.72 & +0.72 \nl
 & & &              $\pm$~ 13 &   12 &   7 &  12 &   8 &  7 &  0.12 &  0.12 \nl
      & & & & & & & & & & \nl
Halo-3    & 162 &--1.61 &  +9 &--210 &--12 & 168 & 102 & 97 & +0.71 & +0.71 \nl
 & & &              $\pm$~ 14 &   12 &   8 &  13 &   8 &  7 &  0.12 &  0.12 \nl
      & & & & & & & & & & \nl
      & & & & & & & & & & \nl
Halo-1R   &  86 &--1.34 &--13 &--216 &--13 & 172 &  92 & 89 & +0.69 & +0.69 \nl
 & & &              $\pm$~ 19 &   16 &  10 &  17 &   9 &  9 &  0.16 &  0.16 \nl
      & & & & & & & & & & \nl
Halo-1P   &  83 &--1.86 & +31 &--195 &--10 & 154 & 111 &100 & +0.73 & +0.73 \nl
 & & &              $\pm$~ 18 &   17 &  12 &  16 &  12 & 10 &  0.18 &  0.18 \nl
      & & & & & & & & & & \nl
      & & & & & & & & & & \nl
Disk-1  &  44 &--0.66 &  +4 & --34 &--18 &  45 &  43 & 25 & +1.24 & +1.08 \nl
 & & &              $\pm$~~ 8 &    8 &   6 &   8 &   8 &  6 &  0.34 &  0.34 \nl
      & & & & & & & & & & \nl
Disk-2    &  38 &--0.58 &  +8 & --43 &--19 &  51 &  47 & 25 & +1.15 & +1.01 \nl
 & & &              $\pm$~~ 9 &   10 &   6 &   9 &   9 &  6 &  0.35 &  0.35 \nl
      & & & & & & & & & & \nl
Disk-3    &  51 &--0.76 &  +6 & --45 &--16 &  52 &  48 & 29 & +0.94 & +0.79 \nl
 & & &              $\pm$~~ 8 &    9 &   6 &   8 &   8 &  5 &  0.30 &  0.30 \nl
      & & & & & & & & & & \nl
All stars & 213 &--1.40 &  +7 &--169 &--14 & 147 & 115 & 85 & +0.73 & +0.73 \nl
 & & &              $\pm$~ 10 &   10 &   6 &  10 &   7 &  6 &  0.11 &  0.11 \nl
\enddata
\end{deluxetable}




\begin{deluxetable}{llccc}
\tablenum{7}
\tablewidth{46pc}
\tablecaption{Selected Absolute Magnitude Determinations, $M_V(RR) = a $[Fe/H]$ + b$. \label{tbl-7}}
\tablehead{
\colhead{Reference} & \colhead{Method} & \colhead{$a$} & \colhead{$b$} & \colhead{Symbol}
}
\startdata
BH86     &  Stat-$\pi$ of 142 field RR Lyraes ($\langle$[Fe/H]$_{L94} \rangle$ = --1.32)\tablenotemark{a} & $0.$            &  $0.79 \pm 0.14$  &  $\pi$ \nl
													                    		      
SRM      &  Stat-$\pi$ of 139 field RR Lyraes ($\langle$[Fe/H]$_{L94} \rangle$ = --1.32)\tablenotemark{a} & $0.$            &  $0.77 \pm 0.14$  &  $\Pi$ \nl
													                    		      
Buonanno {\it et al.} 1990  &  GC main sequence fits to subdwarfs\tablenotemark{b}                        & $0.34 \pm 0.14$ &  $1.1 \pm 0.2$    &  B90   \nl
													                    		      
Jones {\em et al.} 1988     &  MS fit of M5 to best subdwarf ([Fe/H] = --1.4)                             & $0.$            &  $0.86 \pm 0.12$  &  J     \nl
													                    		      
Bolte \& Hogan 1995         &  MS fit of M92 to updated subdwarfs ([Fe/H] = --2.3)                        & $0.$            &  $0.49$           &  B     \nl
													                    		      
CSJ                         &  Synthesis of several $M_V(RR)$ results                                     & $0.15 \pm 0.01$ &  $1.01 \pm 0.08$  &  CSJ \nl
													                    		      
Jones {\it et al.} 1992     &  Baade-Wesselink of 18 field RR Lyraes ($\approx$ CSJ)                      & $0.16 \pm 0.03$ &  $1.02 \pm 0.15$  &  --  \nl
													                    		      
Lee 1990                    &  Synthetic HB theory, $Y_{MS} = 0.23$                                       & $0.17$          & $0.79$            &  L23 \nl
													                    		      
Lee 1990                   &   Synthetic HB theory, $Y_{MS} = 0.20$                                       & $0.19$          & $0.97$            &  L20 \nl
													                    		      
Walker 1992                &  LMC RRs using Cepheid distance scale ([Fe/H] = --1.9)                       & $0.$  & $0.44 \pm 0.11\tablenotemark{c}$  &  W   \nl
													                    		      
%
Gould 1995                  &  LMC RRs using SN1987A ring distance ([Fe/H] = --1.9)                       & $0.$  & $>0.57 \pm 0.06\tablenotemark{c}$ &  $\downarrow$ \nl
													                    		      
Ajhar {\it et al.} 1996    &  M31 cluster HBs using Cephied distance scale                                & $0.08 \pm 0.13$ &  $0.88 \pm 0.21$  &  A96 \nl
													                    		      
Sandage 1990b               &  Sandage period shift effect                                                & $0.39$          &  $1.17 \pm 0.2$   & S90  \nl
													                    		      
Fernley 1993                &  Period shift using ($V$--$K$) colors                                       & $0.19$          &  $0.84$           &  F93 \nl
													                    		      
Sandage 1993                &  Period shift with BFE $T_{eff}$ correction                                 & $0.30$          &  $0.94$           &  S93 \nl

\enddata

\tablenotetext{a}{Corrected to the reddening and apparent magnitude scales
used in this paper, see Sec. 6.}

\tablenotetext{b}{$M_V(ZAHB)$ corrected to $M_V(RR)$ using Eqn. 4 of CSJ:
$V(RR) = V(ZAHB)$ -- 0.05[Fe/H] -- 0.20.}

\tablenotetext{c}{Error estimated from details given in the cited paper.}

\end{deluxetable}

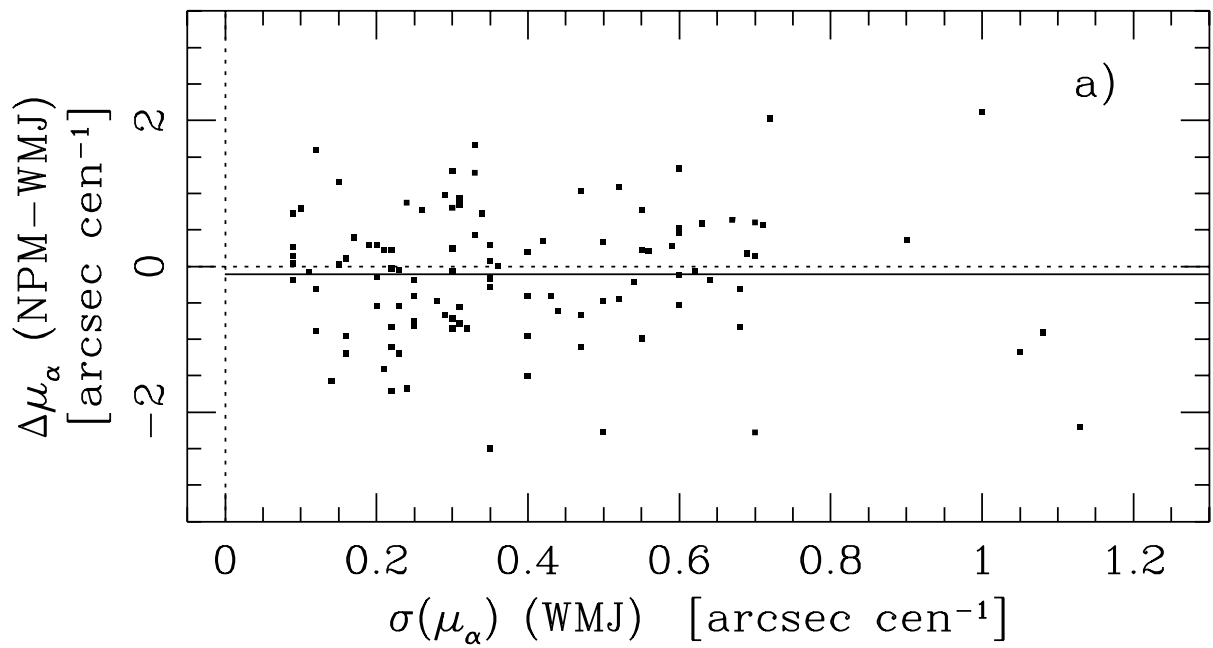
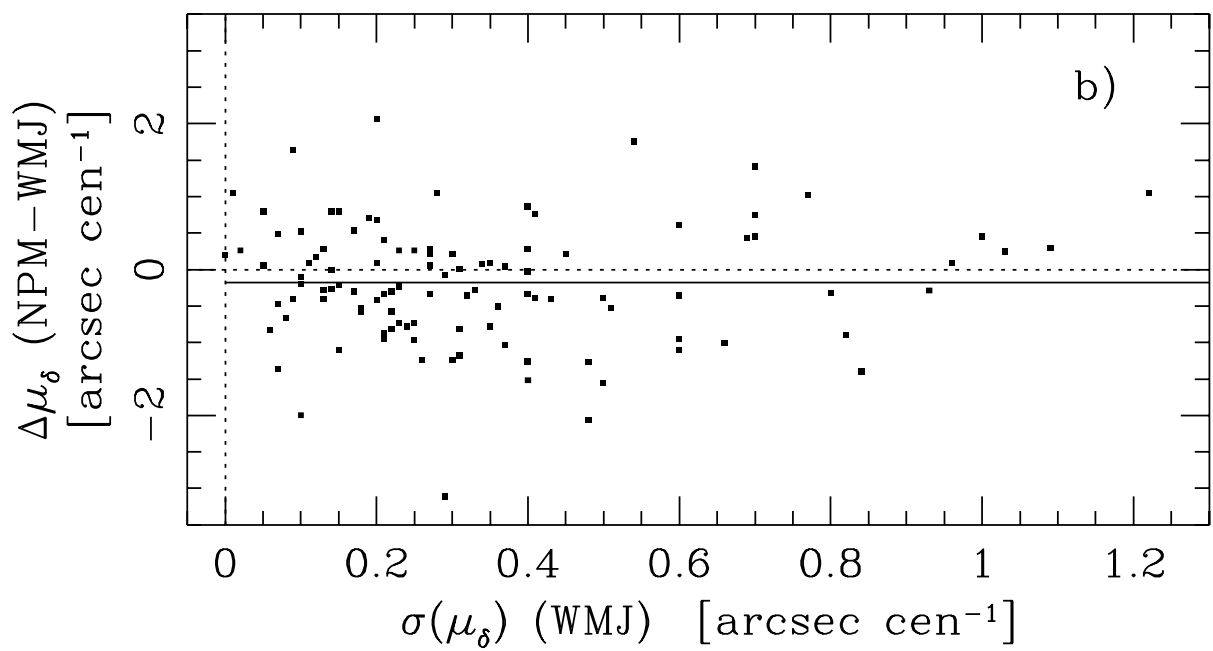
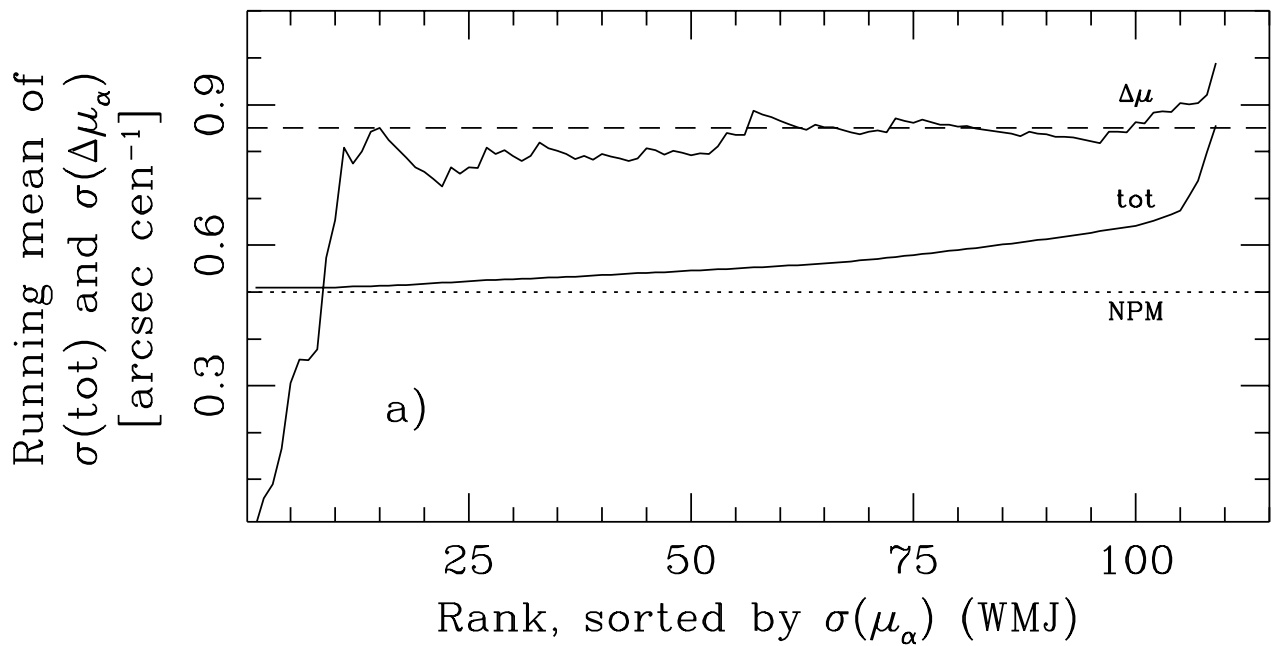
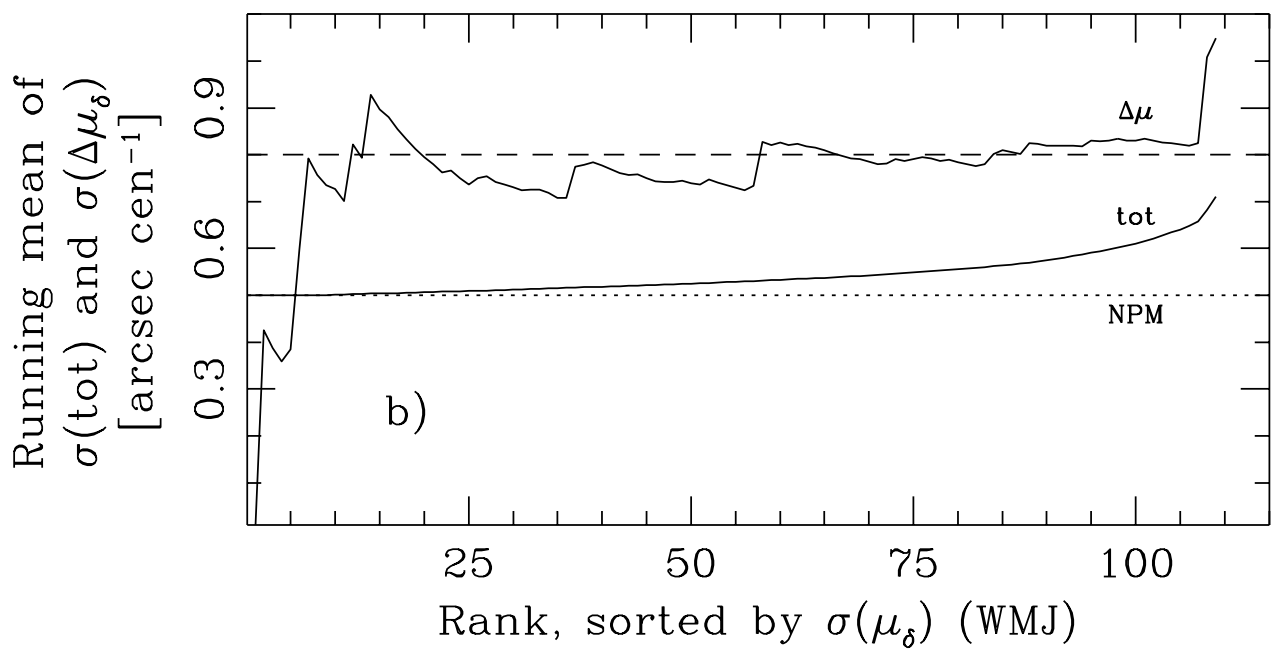
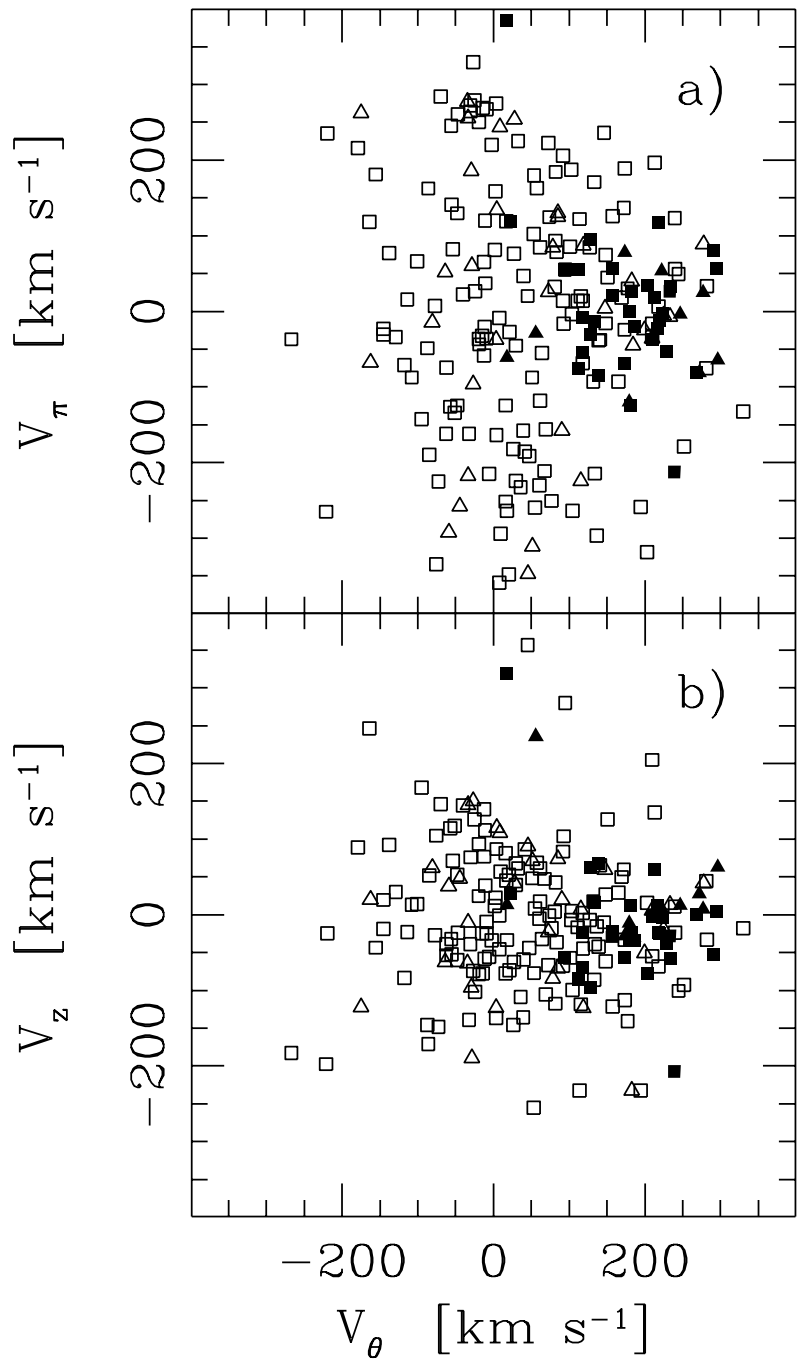
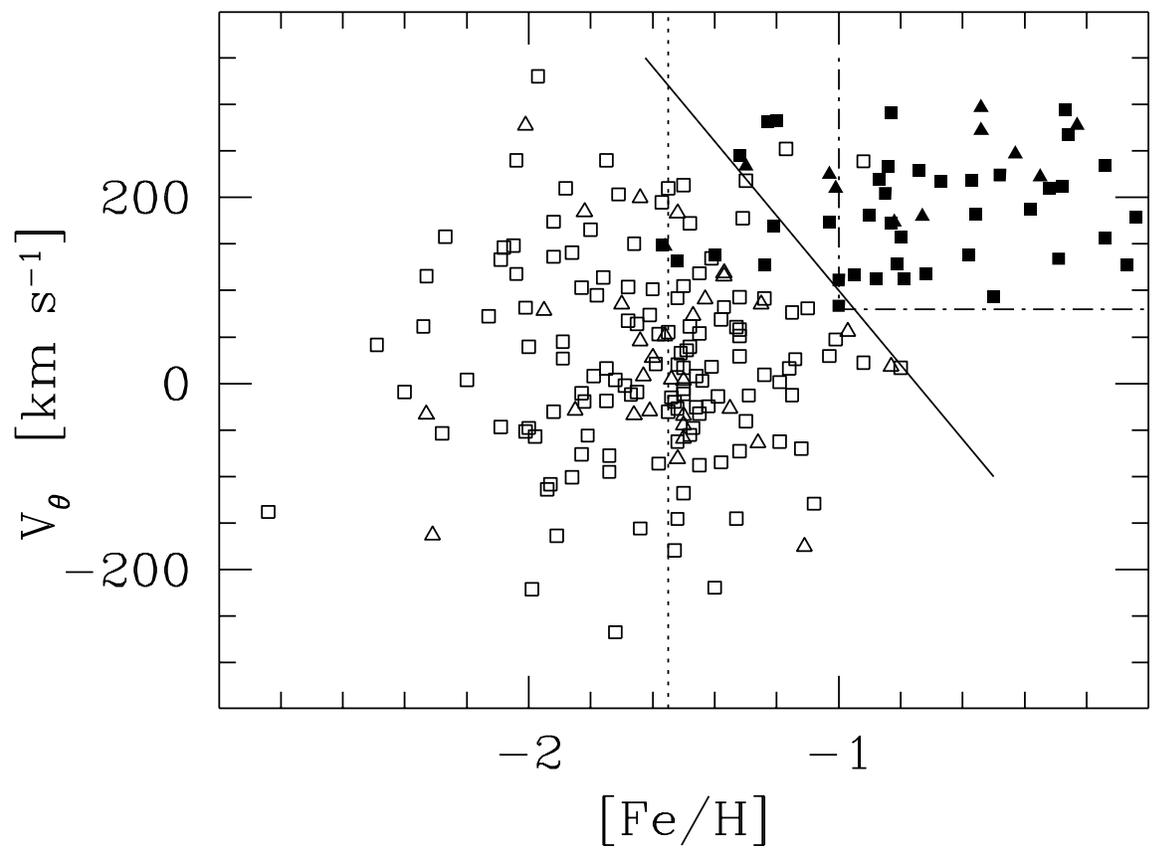
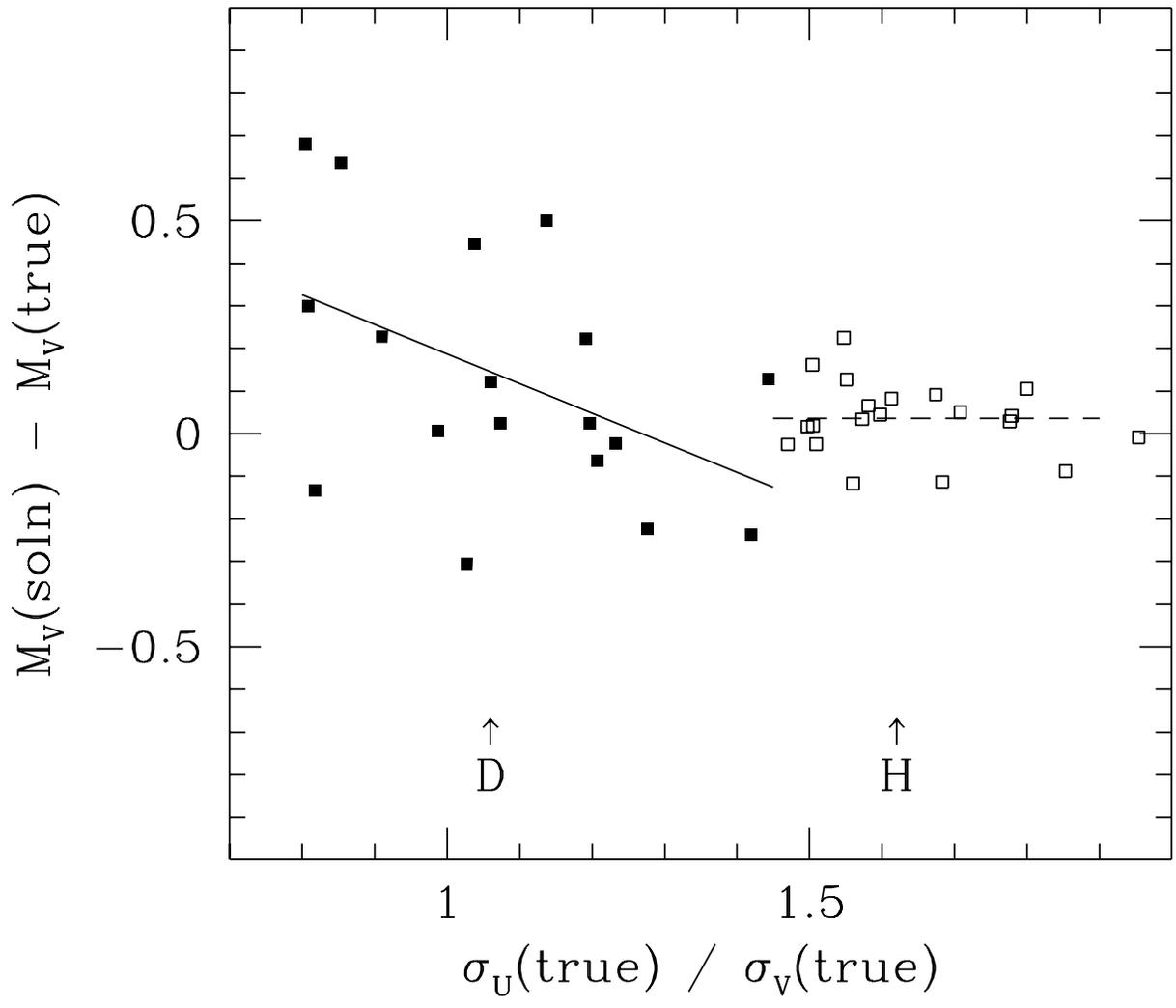
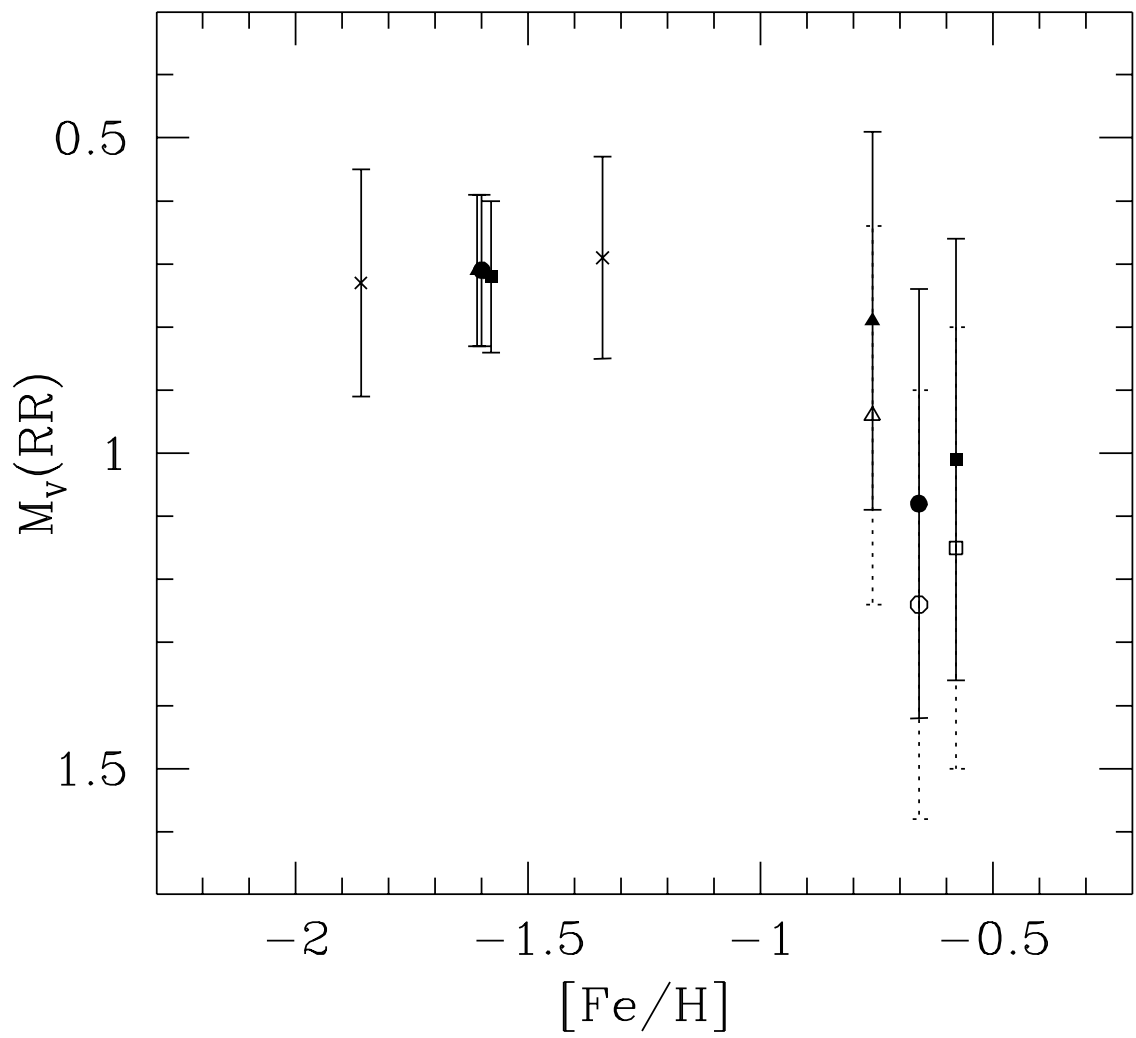
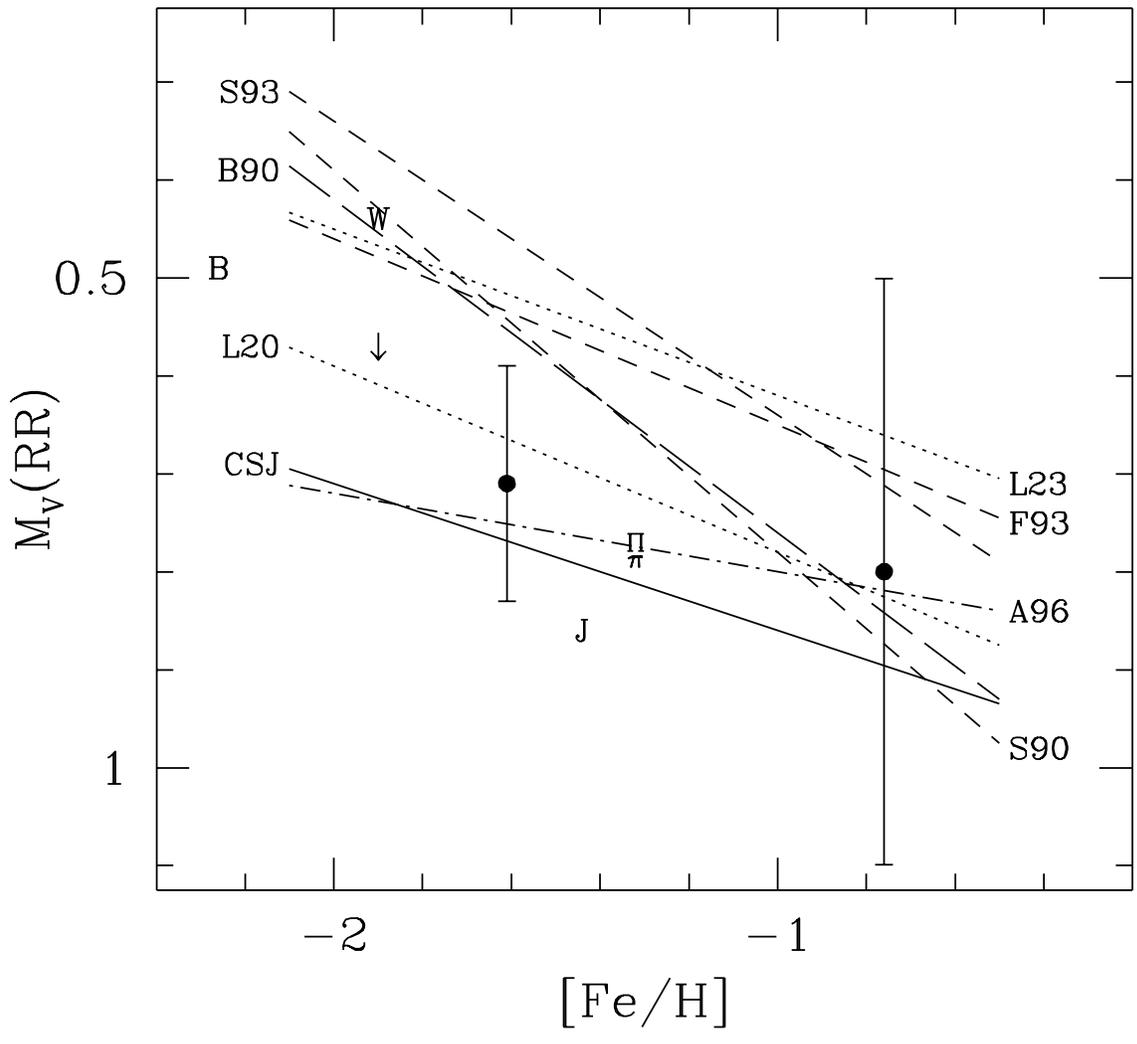
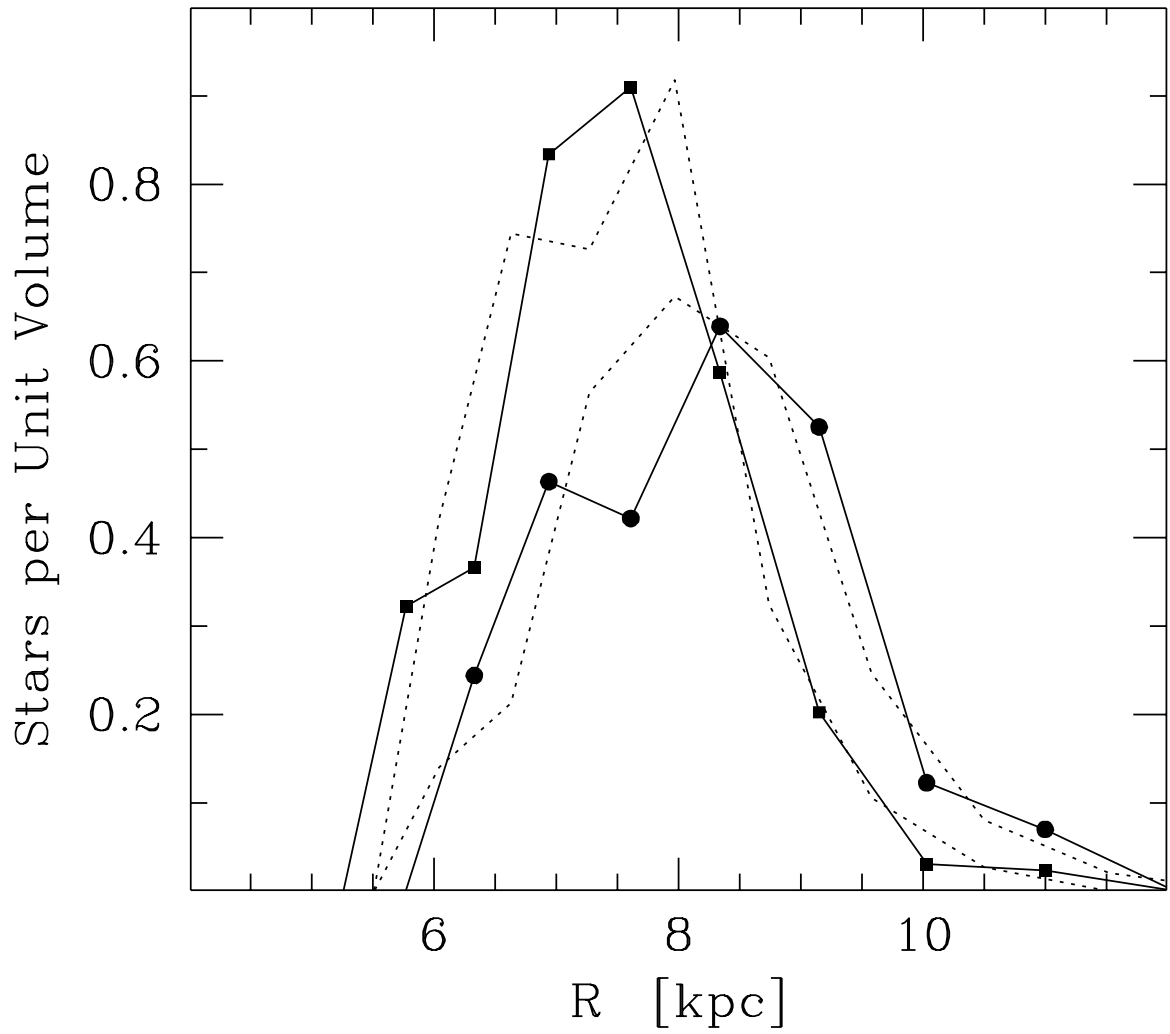
\end{document}